\documentclass[final, 11pt, times, twocolumn]{elsarticle}

\usepackage{amssymb,bm}
\usepackage{amsmath}
\usepackage{upgreek}
\usepackage{booktabs}
\usepackage{float}
\usepackage{esvect}
\usepackage{titlesec}

\usepackage{geometry}
\usepackage[hidelinks]{hyperref}  %
\newcommand{\bmB}{{\bm{B}}}
\newcommand{\bmt}{{\bm{\theta}}}
\newcommand{\rmd}{{\rm{d}}}
\newcommand{\cala}{{\cal A}}
\newcommand{\call}{{\cal L}}
\newcommand{\calz}{{\cal Z}}
\newcommand{\fs}{{\mathcal{F}\text{-statistic}}}
\newcommand{\apc}{{\varnothing}}
\newcommand{\tpc}{{\otimes}}
\usepackage{ulem}
\usepackage{xcolor}
\newcommand{\Refereeone}[1]{#1}
\newcommand{\Refereetwo}[1]{#1}
\newcommand{\ulhref}[2]{\href{#1}{\uline{#2}}}

\journal{SCPMA {\rm as a Short Communication}}

\begin{document}

\begin{frontmatter}

\title{Cracking Gravitational Wave Multiple Ringdown Modes in Space}

\author[inst1,inst2]{Ziming Wang}

\author[inst2]{Han Wang}
\author[inst3,inst4]{Yuxin Yang}

\author[inst1,inst2]{Yiming Dong}

\author[inst5]{Hai-Tian Wang}

\author[inst3,inst4]{Yi-Ming Hu}

\author[inst2,inst6]{Lijing Shao\corref{cor1}}
\ead{lshao@pku.edu.cn}
\cortext[cor1]{Corresponding author.}

\affiliation[inst1]{organization={Department of Astronomy, School of Physics},
            addressline={Peking University},
            city={Beijing},
            postcode={100871},
            country={China}}

\affiliation[inst2]{organization={Kavli Institute for Astronomy and Astrophysics},
            addressline={Peking University},
            city={Beijing},
            postcode={100871},
            country={China}}
\affiliation[inst3]{organization={MOE Key Laboratory of TianQin Mission, TianQin
Research Center for Gravitational Physics, Frontiers Science Center for TianQin,
Gravitational Wave Research Center of CNSA},
            addressline={Sun Yat-sen University (Zhuhai Campus)},
            city={Zhuhai},
            postcode={519082},
            country={China}}

\affiliation[inst4]{organization={School of Physics and Astronomy},
            addressline={Sun Yat-sen University (Zhuhai Campus)},
            city={Zhuhai},
            postcode={519082},
            country={China}}

\affiliation[inst5]{organization={School of Physics},
            addressline={Dalian University of Technology},
            city={Dalian},
            postcode={116024},
            country={China}}

\affiliation[inst6]{organization={National Astronomical Observatories},
            addressline={Chinese Academy of Sciences},
            city={Beijing},
            postcode={100101},
            country={China}}

\begin{abstract}
Ringdown signals from perturbed black holes (BHs) offer a clean window into BH
spacetime, strong-field gravity, and fundamental physics. Presently the
quasi-normal modes of stellar-mass BH ringdowns have been successfully extracted
in the ground-based gravitational wave (GW) observations. Looking ahead, the
future space-borne observatories will listen to the ringdowns from massive BH
binary coalescences more loudly and resolve multiple modes to unprecedented
precision, which calls for efficient approaches to mitigate the sharply
increasing computational burden. We develop a practical ringdown analysis
pipeline for space-borne detectors by implementing FIREFLY, a novel acceleration
algorithm validated in ground-based detectors, and for the first time
demonstrate its compatibility and effectiveness with the time-delay
interferometry (TDI) observables. With high fidelity, we achieve an over two-order-of-magnitude speedup for a simulated ringdown signal including six modes, providing
a viable and scalable route for multi-mode ringdown analysis in the space
context. This new approach has sound statistical interpretation and is
extensible to other GW sources in band.
\end{abstract}

\begin{keyword}
gravitational wave \sep black hole \sep ringdown \sep bayesian analysis
\end{keyword}

\end{frontmatter}

\section{Introduction}

The discovery of gravitational waves (GWs) has opened a new window to explore
our Universe.  Since the first detection of GWs from a binary black hole (BH)
coalescence in 2015~\cite{LIGOScientific:2016aoc}, GW astronomy has flourished
with hundreds of events imposing broad impacts on astrophysics, cosmology, and
fundamental physics~\cite{Bailes:2021tot}. Recently, the LIGO-Virgo-KAGRA (LVK)
Collaboration reported GW250114, the most significant signal to date, and
carried out a dedicated analysis of its ringdown
phase~\cite{LIGOScientific:2025rid}. This epitomizes
the decade-long leap in GW science from first detections to precision
measurements, whereby highlighting the ringdown as a particularly 
incisive probe
in testing general relativity and exploring new physics~\cite{Dreyer:2003bv,Berti:2025hly,LIGOScientific:2025wao,Yang:2025ror}.

Ringdown is the final stage of a binary BH coalescence, where the merger remnant
settles down to a stationary state by emitting GWs. A commonly used description of ringdown is based on the linear BH perturbation
theory, in which the waveform is expanded as a finite sum of damped sinusoids,
namely the QNMs~\cite{Berti:2009kk}.
Measuring the QNM parameters, such as the oscillation frequencies and damping
times---known as the ``BH spectroscopy''---enables unique and powerful studies
of BH spacetime, with the examinations of the
no-hair theorem and Hawking's area law being two representative
examples~\cite{LIGOScientific:2025rid,LIGOScientific:2025wao}.
In practice, it requires extracting
sufficiently rich information from the ringdown. For the first event GW150914,
there has been a long-standing debate regarding the existence of overtones
besides the fundamental mode~\cite{Berti:2025hly,Isi:2019aib,Cotesta:2022pci,Finch:2022ynt}, whereas the overtone was
confidently extracted in GW250114 thanks to its high signal-to-noise ratio
(SNR)~\cite{LIGOScientific:2025wao}.

In the coming decades, the next-generation GW detectors with improved
sensitivity are expected to routinely observe ringdown signals at much higher
SNRs, enabling more precise BH spectroscopy~\cite{Bailes:2021tot,
Bhagwat:2021kwv}. Among them, observations from the massive BH binaries (MBHBs)
in the space-borne detectors, such as the Laser Interferometer Space Antenna
(LISA), TianQin, and Taiji~\cite{Gair:2022knq}, are particularly promising,
where multiple QNMs can be resolved and at least three independent QNM
parameters are to be measured within $1\%$ uncertainty~\cite{Bhagwat:2021kwv,
Shi:2024ttu}. However, extracting multiple QNMs poses challenges on data
analysis. The dimension of the parameter space expands rapidly with the number
of QNMs~\cite{Berti:2009kk}, making the parameter inference more complex and
computationally expensive.  Moreover, in the high-precision regime, to avoid
systematic biases it is necessary to include modes that are not directly
detectable~\cite{Shi:2024ttu, Capuano:2025kkl}, which further complicates the
situation.  Consequently, achieving the promise of BH spectroscopy with
space-borne detectors calls for a fast, accurate, and scalable method for the
increasingly complicated inference problem.

\begin{figure*}[ht]%
\centering%
\includegraphics[width=0.825\textwidth]{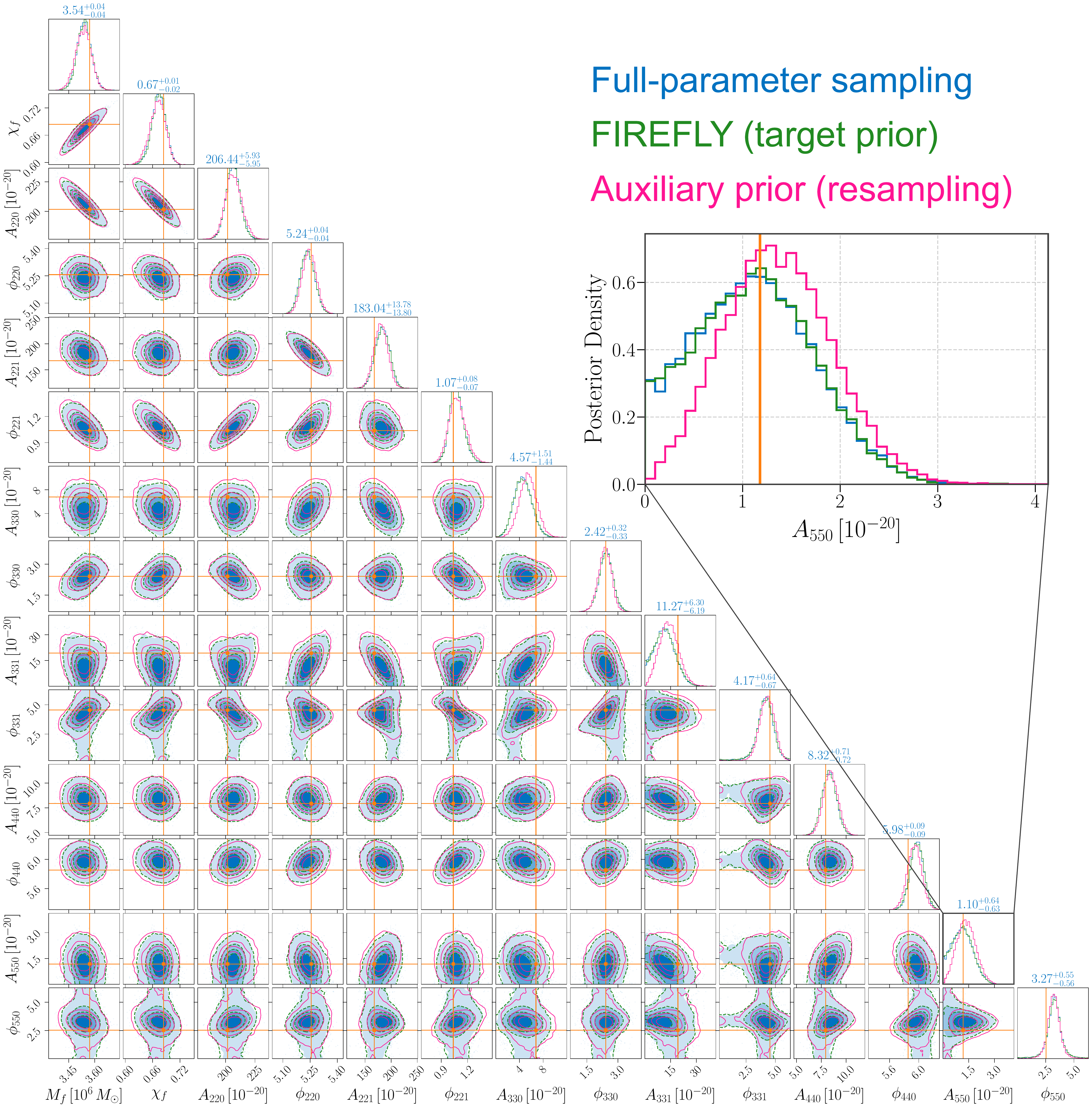}
\caption{Posteriors in the ringdown analysis with six QNMs, where injected values are
indicated by orange lines. Blue, green and pink contours are from the
full-parameter sampling, FIREFLY, and the auxiliary inference, respectively. The
marginal posterior of $A_{550}$ is enlarged for a better inspection.  Contours
are drawn at the two-dimensional 1$\sigma$ (39.3\%), 1.5$\sigma$ (67.5\%),
2$\sigma$ (86.5\%) and 3$\sigma$ (98.9\%) credible levels.
}\label{fig1}
\end{figure*}

In this work, for the first time we design and accelerate a practical ringdown
analysis pipeline for space-borne detectors by implementing
FIREFLY~\cite{Dong:2025igh}, a novel algorithm validated in ground-based
detectors. Based on the mathematical structure of the ringdown signal, FIREFLY
dedicatedly designs Bayesian marginalization and importance sampling to achieve
significant acceleration with clear statistical interpretation and
extensibility.  Here we demonstrate that FIREFLY remains compatible and
effective with the time-delay interferometry (TDI) observables from space-borne
detectors~\cite{Gair:2022knq}.  In simulations including six QNMs,
FIREFLY---with high fidelity---yields an over two orders of magnitude
speedup compared with the
conventional approach.  As the number of QNMs increases, the computational cost
of FIREFLY grows only mildly, positioning itself as a powerful tool for ringdown
analysis. 

\section{FIREFLY for space-borne detectors}

Below we outline the key idea of FIREFLY, and how we incorporate it with TDI
observables.  In  GW data analysis,  the Bayes' theorem is commonly employed to
calculate the posterior $P(x|d)  = P(d|x)P(x)/P(d)$, where $P(x)$ is the prior
representing the knowledge about the parameters $x$ before observing the data
$d$, $P(d|x)$ is the likelihood quantifying the agreement between the data and
model predictions, and $P(d)$ is the evidence serving as a normalization factor.
When the dimensionality of the parameter space is high, it is usually unfeasible
to directly obtain the posterior density, and sampling from the posterior 
become challenging. This reflects the intrinsic difficulty in exploring
high-dimensional parameter spaces, known as the ``curse of dimensionality'',
which cannot be substantially mitigated without extra information.

FIREFLY's acceleration strategy is to avoid direct sampling in the full
parameter space, which is made feasible by recognizing and leveraging the Gaussian structure in the ringdown likelihood.
This feature comes from two facts: (i) the likelihood is quadratic in the
residual between the data $d$ and the model $h$ under Gaussian noise; (ii) the
model $h$, a superposition of QNMs, is linear in the reparametrized mode
amplitudes and phases, denoted by $\bmB$. 
FIREFLY first performs an auxiliary inference that analytically
marginalizes $\bmB$ under a flat prior, sampling only over the other source parameters
$\bmt$ in a low-dimensional space, such as the final BH mass $M_f$ and spin $\chi_f$. Then, the full
posterior of $\bmt$ and $\bmB$ under the target prior is recovered by importance
sampling, where the Gaussian structure again facilitates efficient computation~\cite{Dong:2025igh}.

We now give a schematic derivation for the applicability of FIREFLY to
space-borne detectors, with the detailed derivation given
in the Supplemental Material.  For space-borne detectors, the likelihood is built
from the TDI data channels and remains quadratic in the
residual~\cite{Gair:2022knq}. To
incorporate FIREFLY, the key step is to verify the linearity of the TDI data in
$\bmB$. The waveform of a QNM indexed by $(\ell,m,n)$
 can be expressed as a sum of terms of the form $A_{\ell m n} e^{-i\phi_{\ell m n}} f_{\ell m
n}(t; \bmt)$, where $A_{\ell m n} e^{-i\phi_{\ell m n}}$ is the complex
amplitude, and $f_{\ell m n}(t; \bmt)$ depends on $\bmt$.
Then, the waveform can be written as linear forms in $\bmB$
 under the reparameterization $B^{\ell m n,1}
\equiv A_{\ell m n}\cos\phi_{\ell m n}$ and $B^{\ell m n,2} \equiv A_{\ell m
n}\sin\phi_{\ell m n}$. 
To construct the TDI observables, this waveform is first projected onto the detector in a linear combination of $h_+$ and
$h_\times$, followed by a series of adding and subtracting operations with
appropriate time delays. Since these combinations are linear and depend only on
the detector and source geometry, the TDI observables are linear in $\bmB$,
enabling the use of FIREFLY. However, FIREFLY requires the corresponding linear basis functions to
be explicitly obtained, beyond the standard TDI observable calculation. Thus, the
formal linearity of the TDI observables does not by itself determine the actual
computational speedup, which we assess quantitatively below.

\section{Numerical results}

We adopt the \texttt{GWSpace}~\cite{Li:2023szq} package to simulate TDI signals
and extend it to explicitly construct the linear basis functions
required by FIREFLY.
As an example,
we consider a non-spinning MBHB with source-frame masses $m_{1,\rm s} = 4\times
10^5 M_\odot$ and $m_{2,\rm s} = 3.5\times 10^5 M_\odot$ at redshift $z=4$,
which gives $M_f \approx 3.6\times 10^6 M_\odot$ and $\chi_f \approx 0.68$ in
the detector frame.  The data are recorded from $t = 10 M_f$ after the merger in
the TDI A and E channels of TianQin. \Refereetwo{This choice is for illustration, but the analysis readily
generalizes to BBHs with other masses, other start times $t$, and other detector
(network) configurations. Choosing different source parameters, such as the masses,
changes the signal amplitudes, QNM frequencies, and SNR, but still preserving the
signal's linear dependence on $\bmB$ and hence the applicability of FIREFLY.}  We use uniform prior in $M_f$, $\chi_f$, $A_{\ell m n}$ and
$\phi_{\ell m n}$. More details about the injection and inference are given in
the Supplementary Material.

In Fig.~\ref{fig1}, we show the posterior distributions from the conventional
full-parameter sampling and the FIREFLY, where we include six QNMs with
$(\ell,m,n) = (2,2,0)$, $(2,2,1)$, $(3,3,0)$, $(3,3,1)$, $(4,4,0)$ and
$(5,5,0)$, for a signal with a total SNR $\sim 210$. The two methods give highly consistent results, \Refereeone{while FIREFLY
achieves more than two orders of magnitude speedup, reducing the runtime from
about 13 hours to 4 minutes using \texttt{nessai} with 64 CPU cores, corresponding to approximately 870 and 4.5 allocated
core-hours, respectively.} We also show the
posterior from the auxiliary inference, which adopts a flat prior on $\bmB$ and
(less faithfully) favors larger amplitudes: for the subdominant, low-SNR modes
like $(5,5,0)$, the amplitude from the auxiliary inference is visibly
overestimated. This discrepancy also highlights the importance of a
comprehensive exploration of priors. In our simulations, the importance-sampling step takes only tens of
seconds (see the Supplementary Material), allowing FIREFLY to efficiently repeat
the analysis for different target priors while faithfully recovering the
corresponding posteriors.

\begin{figure}[t]%
\centering%
\includegraphics[width=0.5\textwidth]{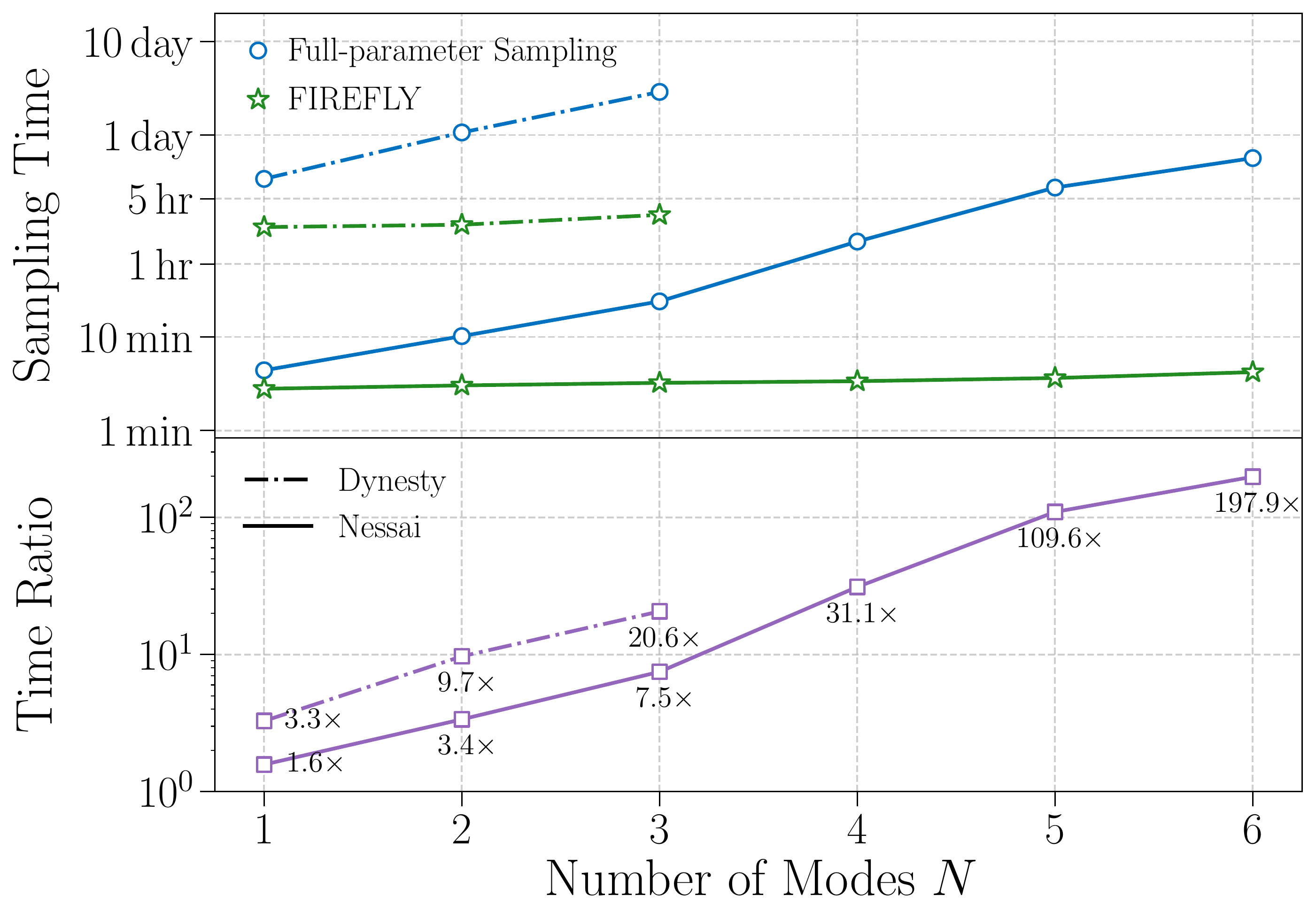}
\caption{Sampling time for full-parameter sampling and FIREFLY, and their ratios
as a function of the number of QNMs, averaged over five independent runs.
Dash-dotted and solid lines show results with \texttt{dynesty} and
\texttt{nessai} samplers, respectively. 
}\label{fig2}
\end{figure}

Figure~\ref{fig2} further shows the time scaling of the two methods with the
number of  QNMs included, with $N$ indicating the first $N$ modes in the above
six QNMs. We adopt two samplers, \texttt{dynesty}~\cite{Higson:2018cwj} and
\texttt{nessai}~\cite{Williams:2021qyt}, for $1\leq N \leq 3$ and $1\leq N \leq
6$, respectively, and report the average time over 5 independent runs. In all
runs, FIREFLY yields consistent results with the full-parameter sampling, with
quantitative metrics provided in the Supplementary Material. The acceleration becomes more
significant for larger $N$, since more parameters are marginalized in the
auxiliary inference step. Additional timing tests with different prior ranges and mode
configurations are presented in the Supplementary Material. Across these tests,
FIREFLY's runtime varies only mildly around a few minutes, demonstrating its
scalability for multimode ringdown analysis in space-borne detectors.
As a more demanding stress test, we further performed an eight-mode
analysis by adding the $(2,1,0)$ and $(2,1,1)$ modes to the six-mode setup. The
full-parameter \texttt{nessai} run did not converge within five days, whereas
FIREFLY produced stable posteriors within a few minutes, consistent with the
injected values and Fisher-matrix predictions; see the Supplementary Material
for details. The computational cost of FIREFLY is also stable across different
noise realizations in the test.

\section{Conclusion and outlook}

In this work, for the first time we validate and implement the novel FIREFLY
algorithm for the ringdown analysis in space-borne detectors, delivering
order-of-magnitude acceleration without compromising accuracy. In this work, the ringdown signal is assumed to be a superposition of linear
QNMs, while
nonlinear effects and non-modal signals may become relevant for high-SNR events.
Nevertheless, the linear parameterization of the
ordinary QNMs can still be exploited for acceleration, and extensions of FIREFLY to
these more general scenarios are probable (Wang et al., in
prep.). In addition,
the framework developed here has the potential to extend to other GW sources in
space-borne detectors' frequency band, such as double white dwarfs and extreme
mass-ratio inspirals, where similar likelihood structures exist. FIREFLY can also be integrated with other acceleration strategies, such as
machine-learning or hardware acceleration.
We expect FIREFLY to ease the computational bottlenecks of future space-borne GW
observations, unlocking their full potential for GW astrophysics.

\medskip\noindent \textbf{Conflict of interest}\par
The authors declare no conflicts of interest.

\medskip\noindent \textbf{Code availability}\par
\Refereeone{The implementation of the linear TDI
    basis functions for FIREFLY is publicly available as part of the official
    \ulhref{https://github.com/TianQinSYSU/GWSpace/releases/tag/v0.2.0}
    {\texttt{GWSpace}} v0.2.0 release. The analysis scripts used in this work are
    available upon reasonable request.}

\medskip\noindent \textbf{Acknowledgements}\par
This work was supported by the National Natural Science Foundation of China
(123B2043, 12573042), the Beijing Natural Science Foundation (1242018), the
National SKA Program of China (2020SKA0120300), the Max Planck Partner Group
Program funded by the Max Planck Society, and the High-performance Computing
Platform of Peking University.

\clearpage
\onecolumn
\appendix

\renewcommand{\thesection}{\Alph{section}}
\renewcommand{\thesubsection}{\thesection.\arabic{subsection}}

\titleformat{\section}
  {\normalfont\bfseries}
  {\thesection.}{0.5em}{}

\titleformat{\subsection}
  {\normalfont\itshape}
  {\thesubsection.}{0.5em}{}

\setcounter{equation}{0}
\renewcommand{\theequation}{\thesection.\arabic{equation}}

\thispagestyle{plain}
\begin{center}
  {\LARGE\bfseries Supplementary Materials}
\end{center}

\section{Linearity of the time-delay interferometry observables in the mode parameters}
\label{app1}

\subsection{Quasi-normal mode waveform and the linear parametrization}

For a given quasi-normal mode (QNM) labelled by non-negative integers
$(\ell,m,n)$, we in fact are concerned with a pair of modes $(\ell,m,n)$ and
$(\ell,-m,n)$, together with their mirror partners~\cite{Berti:2009kk,
Isi:2021iql}. The corresponding complex gravitational wave (GW) strain, $h_{\ell
m n} \equiv h_{\ell m n,+} - i h_{\ell m n,\times}$, is given by
\begin{equation}
    \begin{aligned}
  h_{\ell m n}(t) 
  =
  \left[
    \cala_{\ell m n} e^{- i \Omega_{\ell m n} t}
    + \cala^{\rm mir}_{\ell m n}
      e^{- i \Omega^{\rm mir}_{{\ell m n}} t}
  \right]
  {}_{-2}Y_{\ell m}
  +
  \left[
    \cala_{\ell-mn} e^{- i \Omega_{\ell-mn} t}
    + \cala^{\rm mir}_{\ell-mn}
      e^{- i \Omega^{\rm mir}_{{\ell-mn}} t}
  \right]
  {}_{-2}Y_{\ell-m}\,,
    \end{aligned}\label{eq:full waveform}
\end{equation}
where ${}_{-2}Y_{\ell m}$ are spin-weighted spherical harmonics as  functions of
inclination $\iota$ and azimuth $\varphi$ angles,\footnote{We approximate the
spin-weighted spheroidal harmonics with the spherical harmonics as in most of
ringdown studies, while adopting spheroidal harmonics is straightforward in our
study and does not affect the application of FIREFLY.} and $\cala_{\ell m n}$
are complex amplitudes, $\cala_{\ell m n} = A_{\ell m n} e^{-i\phi_{\ell m n}}$
with real $A_{\ell m n}>0$.  The complex frequencies are written as
$\Omega_{\ell m n} = \omega_{\ell m n} - i/\tau_{\ell m n}$ with $\omega_{\ell m
n}>0$ and $\tau_{\ell m n}>0$.  The mirror frequencies obey $\Omega^{\rm
mir}_{\ell-mn} = - \Omega_{\ell m n}^\ast$, so that $\omega^{\rm mir}_{\ell-mn}
= - \omega_{\ell m n}$ and $\tau^{\rm mir}_{\ell-mn} = \tau_{\ell m n}$. The
complex frequencies are fully determined by the remnant BH mass $M_f$ and spin
$\chi_f$ in general relativity (GR), but they can be treated as free parameters
in model-agnostic analysis. Same as the main text, we use $\bmt$ to denote other
source parameters except for the mode amplitudes and phases.

Assuming mirror symmetry, one has $ \cala^{\rm mir}_{\ell m n} =
(-1)^{\ell}\,\cala^\ast_{\ell-mn}$. We ignore the counter-rotating modes with
$m<0$, as well as the mirror modes of $m>0$.  Now Eq.~\eqref{eq:full waveform}
simplifies to
\begin{equation}
h_{\ell m n}(t)
  =
  {e^{-t/\tau_{\ell m n}}
  \left[
    \cala_{\ell m n}\,{}_{-2}Y_{\ell m} e^{-i\omega_{\ell m n} t}
    + (-1)^{\ell}\cala_{\ell m n}^{\ast}\,{}_{-2}Y_{\ell-m} e^{+i\omega_{\ell m n} t}
  \right]}\,.
\end{equation}
Without these assumptions, the model's
linearity still holds, only involving more linear parameters. Next, we introduce
the combinations
\begin{equation}
  Y^{+}_{\ell m}
  \equiv {}_{-2}Y_{\ell m} + (-1)^{\ell} {}_{-2}Y_{\ell-m}\,,
  \qquad
  Y^{\times}_{\ell m}
  \equiv {}_{-2}Y_{\ell m} - (-1)^{\ell} {}_{-2}Y_{\ell-m}\,,
\end{equation}
and write the polarizations in the damping-oscillator form as
\begin{equation}
    \begin{aligned}
  h_{\ell m n,+}(t)
  &=
  A_{\ell m n}\,Y^{+}_{\ell m}
  \cos \big (\omega_{\ell m n} t + \phi_{\ell m n}\big )
  e^{-t/\tau_{\ell m n}}\,, \\
  h_{\ell m n,\times}(t)
  &=
  A_{\ell m n}\,Y^{\times}_{\ell m}
  \sin \big (\omega_{\ell m n} t + \phi_{\ell m n}\big )
  e^{-t/\tau_{\ell m n}}\,.
    \end{aligned}
\end{equation}
For ringdown signals with multiple modes, the total waveform is obtained by
simply summing over all $(\ell,m,n)$ indices.

FIREFLY works with linear parameters $B^{\ell m n,j}$ ($j=1,2$), instead of the
amplitude and phase $(A_{\ell m n},\phi_{\ell m n})$. Defining
\begin{equation}
  B^{\ell m n,1} \equiv A_{\ell m n}\cos\phi_{\ell m n}\,,
  \qquad
  B^{\ell m n,2} \equiv A_{\ell m n}\sin\phi_{\ell m n}\,,
\end{equation}
and factoring out the purely time-dependent damping cosine and sine terms 
\begin{equation}
  C_{\ell m n}(t) \equiv
  \cos(\omega_{\ell m n} t)\,e^{-t/\tau_{\ell m n}}\,,
  \qquad
  S_{\ell m n}(t) \equiv
  \sin(\omega_{\ell m n} t)\,e^{-t/\tau_{\ell m n}}\,,
\end{equation}
we rewrite the waveform as a linear combination of $B^{\ell m n,j}$,
\begin{equation}
    \begin{aligned}
    h_{\ell m n,+}(t) = &
    \Big[ Y^{+}_{\ell m}\,C_{\ell m n}(t) \Big] B^{\ell m n,1}-
    \Big[ Y^{+}_{\ell m}\,S_{\ell m n}(t) \Big] B^{\ell m n,2}\,, \\
    h_{\ell m n,\times}(t) = &
    \Big[ Y^{\times}_{\ell m}\,S_{\ell m n}(t) \Big] B^{\ell m n,1}+
    \Big[ Y^{\times}_{\ell m}\,C_{\ell m n}(t) \Big] B^{\ell m n,2}\,,
    \end{aligned}\label{eq:linear waveform}
\end{equation}
noting that all terms in square brackets now only depend on $\bmt$ and time $t$,
serving as the basis functions of the waveform.

\subsection{Single link response}

The single link response is the crucial block in building the time-delay
interferometry (TDI) observables.  It describes the relative frequency shift of
the laser light traveling between two spacecrafts, labelled by the ordered pair
of spacecrafts $\rm s \rightarrow \rm r$ and the arm ${\rm l}$.  The positions
of the spacecraft are denoted as $\bm{r}_{\rm s}$ for the sender and
$\bm{r}_{\rm r}$ for the receiver, with the unit vector of the photon
propagation direction given by $\hat{\bm{n}}_{\rm l}$. The direction of GW
propagation is indicated by the unit vector $\hat{\bm{k}}$.  $L_{\rm l}$ is the
length of arm l of the interferometer.  The response due to the GW on the link
${\rm slr}$ can be written as~\cite{Gair:2022knq,Li:2023szq}
\begin{equation}
  y^{\rm slr}(t)
  =
  \frac{
    \hat{n}_{\rm l}^a \hat{n}_{\rm l}^b
  }{
    2\bigl(1 - \hat{\bm{k}}\cdot\hat{\bm{n}}_{\rm l}\bigr)
  }
  \Bigl[
    h_{ab}\bigl(t - L_{\rm l}/c - \hat{\bm{k}}\cdot\bm{r}_{\rm s}/c\bigr)
    - h_{ab}\bigl(t - \hat{\bm{k}}\cdot \bm{r}_{\rm r}/c\bigr)
  \Bigr]\,,
\end{equation}
where $h_{ab}(t)$ represents the metric perturbation with  spatial indices $a$
and $b$, and the Einstein summation convention is adopted.  Next, we express
$h_{ab}(t)$ in terms of the plus and cross polarization modes:
\begin{equation}
  h_{ab}(t)
  = h_+(t)\,e^{+}_{ab} + h_\times(t)\,e^{\times}_{ab}\,,
\end{equation}
where the polarization tensors $e^{+}_{ab}$ and $e^{\times}_{ab}$ depend on the
source's sky location $(\beta,\lambda)$ and polarization angle $\psi$. By
introducing the antenna pattern factors, $ \zeta^{+}_{\rm l} \equiv \hat{n}_{\rm
l}^a \hat{n}_{\rm l}^b e^{+}_{ab}$ and $\zeta^{\times}_{\rm l} \equiv
\hat{n}_{\rm l}^a \hat{n}_{\rm l}^b e^{\times}_{ab}$, the single link response
simplifies to
\begin{equation}
    y^{\rm slr}(t)
    =
    \frac{1}{2\bigl(1 - \hat{\bm{k}}\cdot\hat{\bm{n}}_{\rm l}\bigr)}
    \Bigl[
        H^{\rm l}
        \bigl(
        t - L_{\rm l}/c - \hat{\bm{k}}\cdot\bm{r}_{\rm s}/c
        \bigr)
        -
        H^{\rm l}
        \bigl(
        t - \hat{\bm{k}}\cdot\bm{r}_{\rm r}/c
        \bigr)
    \Bigr]\,,\label{eq:single link response}
\end{equation}
where $H^{\rm l} \equiv \zeta^{+}_{\rm l} h_+ + \zeta^{\times}_{\rm l} h_\times$
and it reflects the projection of the GW onto the link arm ${\rm l}$.

From Eqs.~\eqref{eq:linear waveform} and \eqref{eq:single link response}, the
detector projection and the subsequent time-delay combinations are linear and
independent of $\bmB$.  Therefore, $y^{\rm slr}(t)$ is linear in $\bmB$, and we
explicitly write the basis functions corresponding to $B^{\ell m n,1}$ and
$B^{\ell m n,2}$ as
\begin{equation}
\begin{aligned}
  g^{\rm slr}_{\ell m n,1}(t)
  &= \frac{1}{2\bigl(1 - \hat{\bm{k}}\cdot\hat{\bm{n}}_{\rm l}\bigr)}
  \Bigl\{
    \zeta^{+}_{\rm l} Y^{+}_{\ell m}
      \big[
        C_{\ell m n}(t- L_{\rm l}/c - \hat{\bm{k}}\cdot\bm{r}_{\rm s}/c)
        - C_{\ell m n}(t - \hat{\bm{k}}\cdot\bm{r}_{\rm r}/c)\big] \\
  &\quad\quad\quad\quad\quad\quad\ 
    + \zeta^{\times}_{\rm l} Y^{\times}_{\ell m}
      \big[ S_{\ell m n}(t - L_{\rm l}/c - \hat{\bm{k}}\cdot\bm{r}_{\rm s}/c)
        - S_{\ell m n}(t - \hat{\bm{k}}\cdot\bm{r}_{\rm r}/c)\big]
  \Bigr\}\,, \\
  g^{\rm slr}_{\ell m n,2}(t)
  &= \frac{1}{2\bigl(1 - \hat{\bm{k}}\cdot\hat{\bm{n}}_{\rm l}\bigr)}
  \Bigl\{
    - \zeta^{+}_{\rm l} Y^{+}_{\ell m}
      \big[
        S_{\ell m n}(t - L_{\rm l}/c - \hat{\bm{k}}\cdot\bm{r}_{\rm s}/c)
        - S_{\ell m n}(t - \hat{\bm{k}}\cdot\bm{r}_{\rm r}/c)\big] \\
  &\quad\quad\quad\quad\quad\quad\ 
    + \zeta^{\times}_{\rm l} Y^{\times}_{\ell m}
      \big[
        C_{\ell m n}(t - L_{\rm l}/c - \hat{\bm{k}}\cdot\bm{r}_{\rm s}/c)
        - C_{\ell m n}(t - \hat{\bm{k}}\cdot\bm{r}_{\rm r}/c)\big]
  \Bigr\}\,.
\end{aligned}
\end{equation}
In other words, the single-link response for a single $(\ell,m,n)$ mode now
reads
\begin{equation}
  y^{\rm slr}_{\ell m n}(t)
  =
  \sum_{j=1}^2
  g^{\rm slr}_{\ell m n,j}(t)\,B^{\ell m n,j}\,,\label{eq:linear single link response}
\end{equation}
where we explicitly show the summation over $j$.

\subsection{TDI observables in the linear parametrization}

TDI is a technique to suppress the laser frequency noise in space-borne
detectors.  A generic TDI observable is a linear combination of many single-link
responses with appropriate delay operators.  We index each term in the
combination by $i$, with corresponding label link ${\rm s}_i{\rm l}_i{\rm r}_i$.
Taking the Michelson-X channel as an example, the observable can be formally
written as
\begin{equation}
  h^{X}(t)
  =
  \sum_i
  \mathcal{D}^{X}_{i}
  \bigl[
    y^{{\rm s}_i{\rm l}_i{\rm r}_i}(t)
  \bigr]\,,
\end{equation}
where $\mathcal{D}^{X}_i$ is the delay operator, shifting the argument of time
by $\Delta t^X_i$. The observables in other channels have the same form, only
with different delay operators and single-link responses in the combination. 
For different generation of TDIs, the number of terms in the combination and the
delay time $\Delta t^X_i$ can be different. For example, the first-generation Michelson X combination assumes constant arm lengths and contains eight terms, with delay times given by sums of the relevant arm light-travel times, while its second-generation counterpart accounts for time-varying arm lengths and thus involves more terms and more complicated delay patterns~\cite{Tinto:2020fcc}.
Notably for our study, the
time delay $\Delta t^X_i$ only depends on the detector geometry and source sky
location, but not on the mode parameters $\bmB$. Therefore, we conclude that the
TDI observable $h^{X}(t)$ is linear in $\bmB$. Formally, we substitute
Eq.~\eqref{eq:linear single link response} into the TDI combination and obtain
\begin{equation}
    h^{X}_{\ell m n}(t)
  =
  \sum_i
  \mathcal{D}^{X}_i
  \bigl[
    y^{{\rm s}_i{\rm l}_i{\rm r}_i}_{\ell m n}(t)
  \bigr]
  =
  \sum_{j=1}^{2}
  \Biggl[
    \sum_i
    \mathcal{D}^{X}_i
    g^{{\rm s}_i{\rm l}_i{\rm r}_i}_{\ell m n,j}(t)
  \Biggr]
  B^{\ell m n,j}
  =
  \sum_{j=1}^{2}
  G^{X}_{\ell m n,j}(t)\,B^{\ell m n,j}\,,
\end{equation}
where $G^{X}_{\ell m n,j}(t)$ represents the basis function for $B^{\ell m n,j}$
in the TDI channel $X$, 
\begin{equation}
  G^{X}_{\ell m n,j}(t)
  \equiv
  \sum_i
  \mathcal{D}^{X}_i
  g^{{\rm s}_i{\rm l}_i{\rm r}_i}_{\ell m n,j}(t)\,.\label{eq:basis function in TDI}
\end{equation}
Clearly, this can be understood as directly applying the TDI combination to the
basis functions in the single-link response.

Finally, summing over all QNMs yields the recorded ringdown signal in TDI
channel $X$,
\begin{equation}
  h^{X}(t)
  =
  \sum_{\ell,m,n,j}
  G^{X}_{\ell m n,j}(t)\,B^{\ell m n,j}\,.
\end{equation}
This is exactly the linear structure required in the FIREFLY algorithm, where
$\bmB$ encodes the amplitude and phase of each QNM, and the basis functions
$G^{X}_{\ell m n,j}(t)$ depend on other ringdown parameters $\bmt = \big\{M_f,
\chi_f,\cdots\big\}$. 
In our implementation, we extend the \texttt{GWSpace} package~\cite{Li:2023szq}
to explicitly construct these basis functions, which are not returned by a standard TDI
waveform call. We also pre-compute and store reusable quantities, such as the spacecraft
orbits, to avoid redundant calculations.

\section{FIREFLY algorithm and its implementation in space-borne detectors}

Here, we revisit the FIREFLY algorithm~\cite{Dong:2025igh} and develop its
reformulation for space-borne detectors, based on the linear structure of TDI
observables in the previous section. Before the derivation, we introduce some
notation in the Bayes' theorem
\begin{equation}
    P(x| d) = \frac{ P(d |x)P(x)}{P(d)} = \frac{\call(x)\pi(x)}{\calz}\,,
    \label{eq:Bayes theorem}
\end{equation}
where $\pi(x)\equiv P(x)$ is the prior, $\call(x) \equiv P(d|x)$ is the
likelihood, and $\calz \equiv P(d)$ is the evidence. We use $\tpc$ and $\apc$ to
denote the target posterior and the auxiliary posterior, respectively. Here, $x$
represents the full set of ringdown parameters (decomposed into $\bmt$ and
$\bmB$), while $d$ represents  data from  space-borne detectors. We use Greek
letters to unify the upper indices of $B^{\ell m n,j}$, such that $B^{\mu}$ with
$\mu$ running from 1 to $2N_{\rm mode}$.

We start from the standard likelihood in the GW data analysis assuming Gaussian
noise with zero mean, where the log-likelihood
reads~\cite{Isi:2021iql,Gair:2022knq}
\begin{equation}
    \ln \call=-\frac{1}{2}\langle d - h | d - h \rangle +C_0\,,\label{eq:GW orignal likelihood}
\end{equation}
where $C_0$ does not affect our derivation and is fixed to zero. The inner
product between two signals, $\langle \cdot | \cdot \rangle$, is defined by, 
\begin{equation}
    \langle {h}_1\vert
    {h}_2\rangle=\vv{h}_1^{\intercal}\mathcal{C}^{-1}\vv{h}_2\,,
\end{equation}
where we calculate the inner product in the time domain. The arrow notation
$\vv{h}$ indicates that it is a discrete time series sampled from the continuous
signal $h(t)$, and $\mathcal{C}^{-1}$ is the inverse of the noise covariance
matrix, which is determined by the noise's auto-correlation function, or
equivalently, the power spectral density. In the context of space-borne
detectors, $h$ and $d$ are understood as the model and data in one TDI channel
with corresponding noise properties. For multiple detectors with multiple TDI
channels, one only needs to modify the inner product by summing over all
channels\footnote{For correlated channels such as XYZ channels, the likelihood
is not the product of individual channels, but still has a quadratic form in the
residual $d-h$ with the inner product defined by the full covariance matrix of
all channels.  } and detectors, and the log-likelihood still has a quadratic
form in the residual $d-h$. For brevity, in the following we only consider a
single channel and omit the channel index. 

Next, we rearrange the likelihood in Eq.~\eqref{eq:GW orignal likelihood} to
explicitly shows its Gaussian structure in $\bmB$, giving
\begin{equation}
    \ln \mathcal{L} = {\cal F} - \frac{1}{2} \Big[\big(B^{\mu}-
    \hat{B}^{\mu}\big) M_{\mu\nu} \big(B^{\nu}- \hat{B}^{\nu}\big)  + \langle d
    | d \rangle \Big] \,,
    \label{eq:ll2}
\end{equation}
where $M_{\mu\nu} = \langle G_\mu | G_\nu \rangle$ and $\hat{B}^{\mu} =
(M^{-1})^{\mu\nu} s_{\nu}$ with $s_{\nu} = \langle d | G_\nu \rangle$ and
$G_\nu$ given in Eq.~\eqref{eq:basis function in TDI}. The first term, ${\cal
F}= \frac{1}{2} s_{\mu} (M^{-1})^{\mu\nu} s_{\nu}$, is called the
$\fs$~\cite{Wang:2024jlz}. Note $\hat{B}^{\mu}$, $M_{\mu\nu}$ and $\cal F$ only
depend on $\bmt$ and the data $d$,  not on $\bmB$. 

In the auxiliary inference step, FIREFLY adopts a flat prior on $\bmB$, which is
independent of $\bmt$, while keeping the same prior on $\bmt$ as in the target
inference. Then, the auxiliary posterior is given by
\begin{equation}
    P(\bmt, \bmB | d,\apc) \propto \call(\bmB, \bmt) \pi(\bmt)\pi(\bmB|\apc) \propto
     \exp\Bigl[{\cal F} - \frac{1}{2} \bigl(B^{\mu}- \hat{B}^{\mu}\bigr) M_{\mu\nu} \bigl(B^{\nu}- \hat{B}^{\nu}\bigr)\Bigr] \pi(\bmt)\pi_\bmB\,,
\end{equation}
where $\pi_\bmB = \pi(\bmB|\apc) $ is a constant. Marginalizing over $\bmB$
gives the auxiliary posterior of $\bmt$,
\begin{equation}
    P(\bmt | d,\apc) \propto \sqrt{{\rm det}\big(M^{-1}\big) }e^{\cal F} \pi(\bmt)\,,
\end{equation}
and the stochastic sampling algorithm is performed only in the $\bmt$ space,
drawing samples $\{\bmt_i|\apc\}$. To complete the auxiliary inference, we also
need to sample $\bmB$, whose conditional posterior given $\bmt$ is Gaussian,
 \begin{equation}
     \begin{aligned}
	 P(\bmB | \bmt, d,\apc)
	 =\frac{
	 \call(\bmt,\bmB)\pi(\bmB|\bmt,\varnothing)}{
       \int \call(\bmt,\bmB) \pi(\bmB|\bmt,\varnothing) \rmd\bmB 
     }=\frac{\exp\left
	 \{-\frac{1}{2}\big(B^{\mu}- \hat{B}^{\mu}\big) M_{\mu\nu}
	 \big(B^{\nu}- \hat{B}^{\nu}\big)\right \}}{(2\uppi)^{N}\sqrt{{\rm
	 det}\big(M^{-1}\big)}}\,.
     \end{aligned}\label{eq:conditional posterior of B}
 \end{equation}
Therefore, for each sample $\bmt_i$, the corresponding $\bmB_i$ is generated by
drawing once from the Gaussian distribution ${\cal N}\Big(\hat{\bmB}(\bmt_i),
M^{-1}(\bmt_i)\Big)$. To calculate the full target posterior, we only need
$\{\bmt_i|\apc\}$ instead of $\{(\bmt_i, \bmB_i)|\apc\}$. Here we calculate the
full auxiliary posterior samples for a better comparison with the target
posterior, showing the importance of free prior choices in the GW ringdown
analysis.

In the importance sampling step, FIREFLY adopts a two-step importance sampling
to recover $\bmt$ and $\bmB$, respectively~\cite{Dong:2025igh}. The importance
weight for $\bmt$ is given by 
\begin{equation}
        w^m(\bmt) = \frac{P(\bmt|d,\tpc)}{P(\bmt|d,\apc)}\propto
    \frac{\pi(\bmt)}{\pi(\bmt)}
    \frac{\int \call(\bmt,\bmB) \pi(\bmB|\bmt,\tpc) \rmd\bmB}{\int \call(\bmt,\bmB) \pi(\bmB|\bmt,\apc) \rmd\bmB}\propto 
    \frac{\int \pi(\bmB|\bmt,\tpc) e^{-\frac{1}{2}(B^{\mu}-
  \hat{B}^{\mu}) M_{\mu\nu} (B^{\nu}- \hat{B}^{\nu})} \rmd \bmB}{\sqrt{{\rm
  det}\big(M^{-1}\big) }}\,.
\end{equation}
Note that the integral in the numerator can be efficiently evaluated by an Monte
Carlo (MC)
integration with the Gaussian proposer ${\cal N}\big(\hat{\bmB}(\bmt_i),
M^{-1}(\bmt_i)\big)$, to draw $n_\tpc$ reweighted samples $\{\bmt_j|\tpc\}$.  In
the second step, samples of $\bmB$ are drawn from the conditional posterior
$P(\bmB|d,\bmt,\tpc)$,
\begin{equation}
    P(\bmB | \bmt, d,\tpc)\propto \pi(\bmB|\bmt,\tpc)e^{-\frac{1}{2}(B^{\mu}-
    \hat{B}^{\mu}) M_{\mu\nu} (B^{\nu}- \hat{B}^{\nu})}\,,
\end{equation}
which can also be performed by the importance-sampling technique. For each
$\bmt_j \in \{\bmt_j|\tpc\}$, FIREFLY draws $n_{\rm B}$ samples $\{\bmB_{k|j}\}$
from ${\cal N}\big(\hat{\bmB}(\bmt_j), M^{-1}(\bmt_j)\big)$, then calculates the
weight $w_{k|j} = \pi(\bmB_{k|j}|\bmt_j,\tpc)$. The resampled $\bmB_j$ assigned
to $\bmt_j$ is generated by a single draw from the weighted sample set
$\big\{\bmB_{k|j}, w_{k|j} \big\}$. After repeating this procedure for all
$\bmt_j \in \{\bmt_j|\tpc\}$, $n_\tpc$ samples---denoted as
$\big\{\bmt_j,\bmB_j|\tpc\big\}$---are obtained, which are regarded as being
drawn from the posterior distribution $P(\bmt,\bmB|d,\tpc)$. 

In addition to the posterior distribution, FIREFLY can also calculate the
evidence $\calz$, by appropriately considering the various constant factors in
formulae above. Compared with the posteriors, the evidences given by FIREFLY and
the full-parameter sampling have slight differences at the level of $10^{-3}$,
which increase with the number of modes included in the
analysis~\cite{Dong:2025igh}. One possible reason is the insufficient
exploration of the $\bmB$ parameter space in the full-parameter sampling, which
is more severe for more modes. Though including the evidence calculation is
straightforward, in this study we conservatively apply the FIREFLY for only
calculating the posteriors. A detailed comparison and explanation about the
evidence is left for future work.

\section{Injection and inference}

In the main text, we present an example of a non-spinning MBHB with source-frame
masses $m_{1,\rm s} = 4\times 10^5 M_\odot$ and $m_{2,\rm s} = 3.5\times 10^5
M_\odot$ at redshift $z=4$, giving $M_f \approx 3.6\times 10^6 M_\odot$ and
$\chi_f \approx 0.68$ in the detector frame.  The inclination angle is set to
$\iota = \pi/4$, the polarization angle is set to $\psi = \pi/3$, and the
ecliptic longitude and latitude are set to $\lambda = \pi/5$ and $\beta =
-\pi/6$, respectively. The QNM amplitudes and phases are calculated based on
\texttt{Jaxqualin}~\cite{Cheung:2023vki} package, a package extracting quasinormal modes from BH ringdown simulations, shown in
Table~\ref{tab:qnm_injection_tdiA}. \Refereeone{We generate the detector response using the
analytic TianQin orbit model implemented in \texttt{GWSpace}, treating the
constellation as a rigid equilateral triangle with a constant arm length
$L=\sqrt{3}\times10^{8}\,\mathrm{m}$. We use the first-generation TDI $A$ and
$E$ observables and record the data from $t=10M_f$ after the merger, with a
sampling rate of $1\,\mathrm{Hz}$ and a duration of $5000\,\mathrm{s}$.} The signal-to-noise ratio (SNR) for all
six modes summed is lower than the SNR of the dominant mode $(2,2,0)$, since the
signals of different modes are not orthogonal to each other.

\begin{table}[t]
\centering
\caption{The injected amplitudes $A_{\ell m n}$, phases $\phi_{\ell m n}$ at $t
= 10 M_f$ after the merger, and the individual SNRs which are calculated in the
A and E channels of TianQin, also starting from $t = 10 M_f$.
}
\label{tab:qnm_injection_tdiA}
\vspace{0.2cm}
\setlength{\tabcolsep}{10pt}
\begin{tabular}{@{}cccc@{}}
\toprule
$(\ell,m,n)$ & $A_{\ell m n}\,[10^{-20}]$ & $\phi_{\ell m n}$ & $\rm SNR$ \\
\midrule
$(2,2,0)$ & $202$   & $5.3$ & $279$ \\
$(2,2,1)$ & $168$   & $1.0$ & $168$ \\
$(3,3,0)$ & $6.8$   & $2.4$ & $17.9$ \\
$(3,3,1)$ & $19$    & $4.5$ & $33.8$ \\
$(4,4,0)$ & $7.9$   & $5.9$ & $23.5$ \\
$(5,5,0)$ & $1.2$   & $2.5$ & $3.07$ \\
\midrule
All 6 modes & --- & --- & $212$ \\
\bottomrule
\end{tabular}
\end{table}

In the inference, we adopt the same flat prior for $\bmt$ in both the target and
auxiliary inference. The prior ranges are determined with the Fisher information
matrix~\cite{Vallisneri:2007ev}, and chosen to be 10 times of the standard
deviation in the Fisher matrix estimation for each parameter, which are wide
enough to cover the posterior in practice. For the QNM amplitudes and phases,we
choose flat priors: $\phi_{\ell mn}\sim {\cal U}(0,2\uppi)$ and $A_{\ell mn}\sim
{\cal U}(0,A_{\max})$ with $A_{\max}=500\times 10^{-20}$. In the auxiliary
inference, the flat prior in $\bmB$ corresponds to $\pi\big(A_{\ell mn},
\phi_{\ell mn}\big|\bmt,\apc\big) \propto A_{\ell mn}$. We adopt this prior
shape for all $A_{\ell mn}$ and $\phi_{\ell mn}$, and normalize it in the region
of $A_{\ell mn}\leq A_{\max}$ and $0\leq \phi_{\ell mn}<2\uppi$.  This prior
range is large enough to make the analytical marginalization over $\bmB$ in the
auxiliary inference applicable. Other geometric parameters such as the sky
location are fixed to their injected values, but we can also include them in
$\bmt$ and sample in the inference. In addition, changing the injected geometric
parameters does not affect the speedup substantially.

We adopt the \texttt{bilby}~\cite{Ashton:2018jfp} implementation of two
samplers, \texttt{dynesty}~\cite{Higson:2018cwj} and
\texttt{nessai}~\cite{Williams:2021qyt}, to draw posterior samples for both the
auxiliary inference in FIREFLY and the full-parameter sampling. All samplers are
configured with ${\rm nlive =3000}$ live points and a termination criterion of
${\rm dlogz=0.01}$. We evaluate the MC integration of the first
importance-sampling step with $n_{w}=50000$ draws. We draw the reweighted
samples $\{\bmt_j|\tpc\}$ from the auxiliary posterior with $n_\tpc=20000$
samples, and the number of samples drawn from the conditional posterior of
$\bmB$ is set to $n_{\rm B}=5000$ for each $\bmt_j \in \{\bmt_j|\tpc\}$. In all
samplings, we utilize 64 CPU cores for computation on the FusionServer X6000 V6
equipped with two Intel Xeon Platinum 8358 processors and 256\,GB of RAM.

\section{Quantitative metrics of FIREFLY posterior fidelity and runtime}
\label{sec:Quantitative metrics}

In the main text, we present the posteriors including all six QNMs in Fig. 1 and
find that the results from FIREFLY and the full-parameter sampling are visually
in excellent agreements. In Fig. 2 of the main text, we show the time scaling of
the two methods with the number of included QNMs, denoted as $N$. For each $N$,
we perform 5 independent runs. Here, we further validate the consistency between
the two methods by employing more quantitative measures.

\begin{figure}[H]
    \centering
    \includegraphics[width=0.9\textwidth]{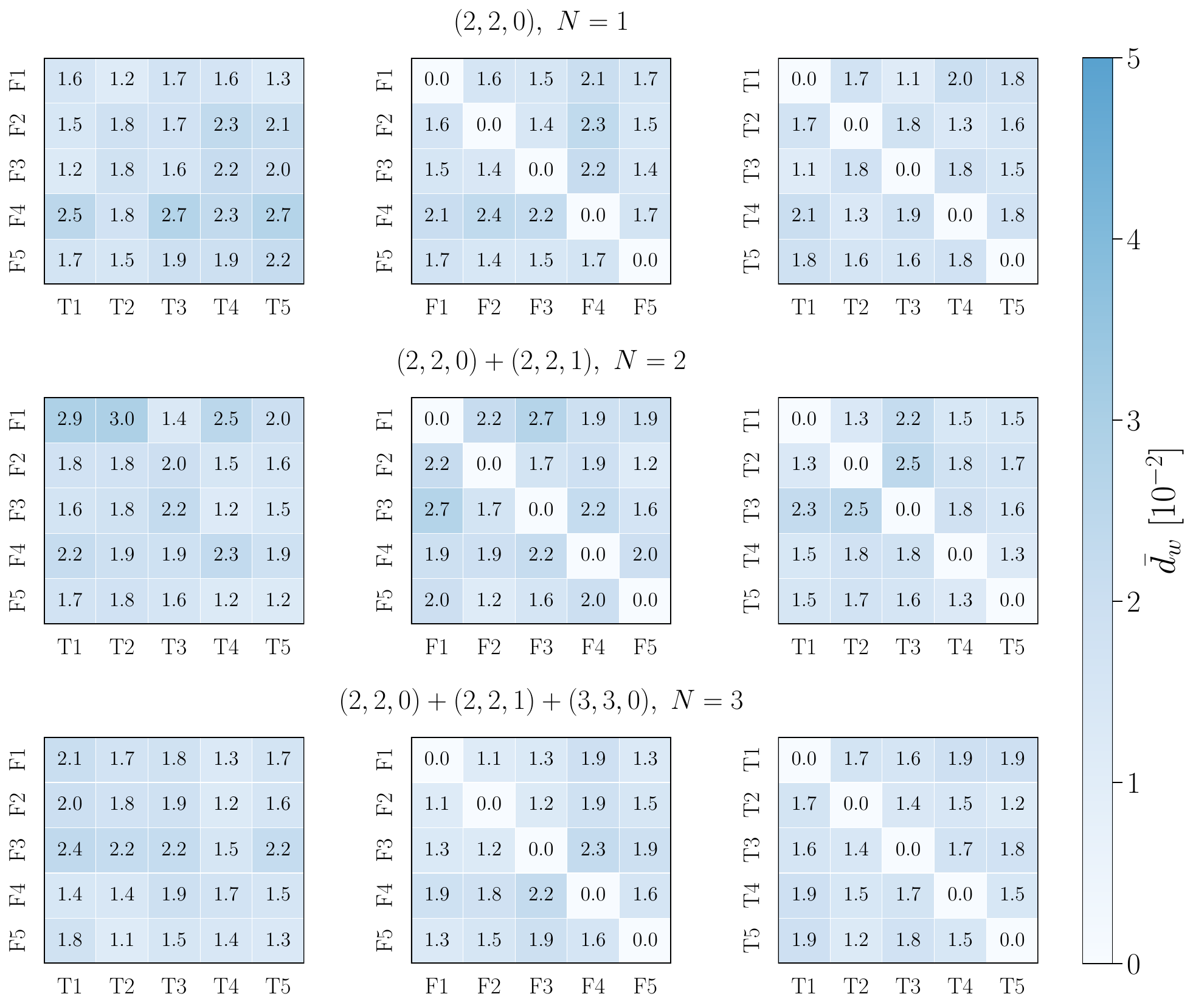}
    \caption{Averaged and normalized one-dimensional Wasserstein distances for
    the posteriors of two methods for $N=1,2,3$ with \texttt{nessai} sampler.
    In each row, the three subplots from left to right show the comparisons
    between FIREFLY and the full-parameter sampling, the comparisons within
    FIREFLY, and the comparisons within the full-parameter sampling. In each
    subplot, the rows labeled from ``F1'' to ``F5'' are five independent
    inferences using FIREFLY, while the columns labeled from ``T1'' to ``T5''
    are those in the full-parameter sampling. Each element in the matrix
    represents the averaged and normalized Wasserstein distance between the
    one-dimensional marginal posteriors.}
\label{fig:WD123}
\end{figure}

\begin{figure}[H]
    \centering
    \includegraphics[width=0.9\textwidth]{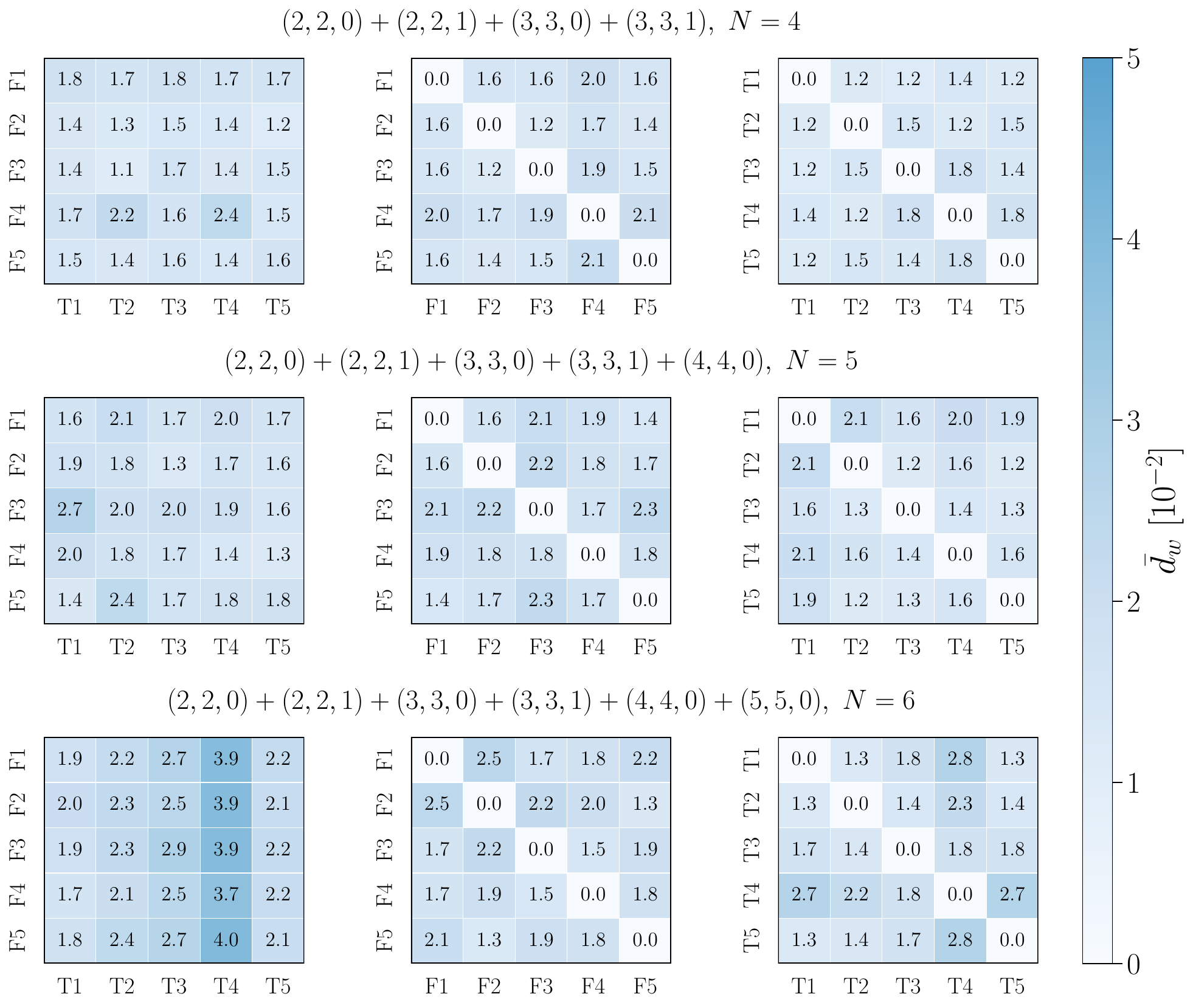}
    \caption{Same as Fig.~\ref{fig:WD123}, but for the cases with $N=4,5,6$ modes.}
\label{fig:WD456}
\end{figure}

\begin{figure}[H]
    \centering
    \includegraphics[width=0.9\textwidth]{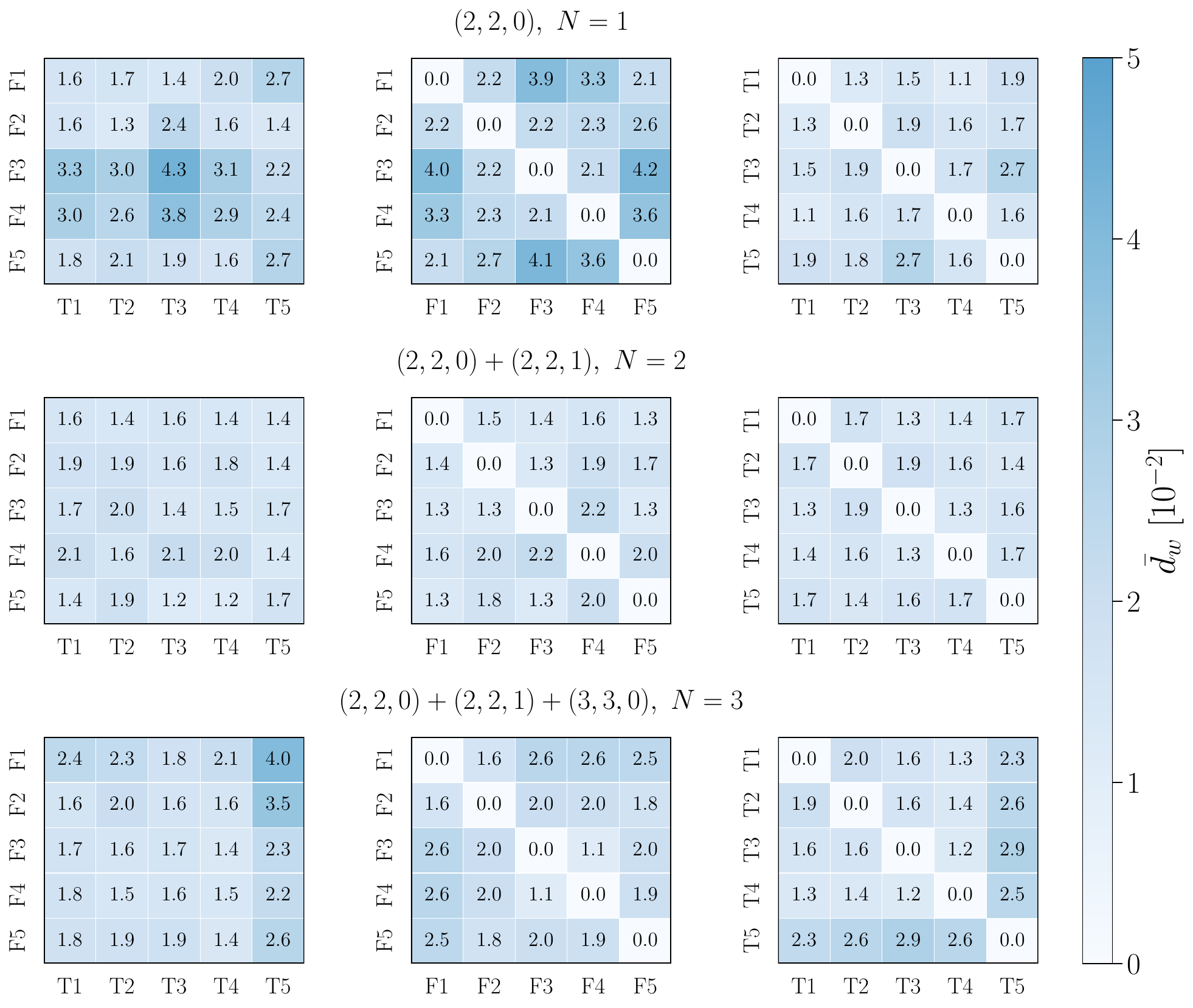}
    \caption{Same as Fig.~\ref{fig:WD123}, but for \texttt{dynesty} sampler.}
\label{fig:WD123d} 
\end{figure}

We use the Wasserstein distance~\cite{MAL-073}, also known as the Earth-Mover's
distance, to quantitatively compare two distributions, denoted as A and B.
First, we compute the one-dimensional Wasserstein distance between each pair of
the one-dimensional marginal posteriors from A and B. Then, we normalize this
distance by the standard deviation of the one-dimensional posterior distribution
in B. The normalized Wasserstein distances are further averaged over all
parameters, providing a single, average, normalized Wasserstein distance,
$\bar{d}_w$.  Here B is the posterior from the full-parameter sampling.  For
reference, we also compute the average normalized Wasserstein distance between
different runs of the same method to reflect its internal variability. In this
case, we treat the row posterior as A and the column posterior as B.
Normalization still uses the standard deviations from the full-parameter
run with the corresponding column index. These
results are shown in Figs.~\ref{fig:WD123} and \ref{fig:WD456} for the sampler
\texttt{nessai}, and in Fig.~\ref{fig:WD123d} for the sampler \texttt{dynesty}.
In all cases, the distances between the two methods are at the same level as the
internal fluctuations within each method and are less than $5\%$ of the standard
deviation, demonstrating the fidelity of FIREFLY in recovering the posteriors
from the full-parameter sampling.







In Table~\ref{tab:firefly_runtime_modes}, we list the stepwise runtime of
FIREFLY corresponding to the reports in Fig. 2 of the main text, for both the
\texttt{nessai} and \texttt{dynesty} samplers. The total runtime of FIREFLY with \texttt{nessai}.
remains at the level of a few minutes and increases only mildly with $N$. The
main computational cost comes from the
auxiliary-inference step, while the importance-sampling step only takes
tens of seconds. Therefore, once
the auxiliary samples are obtained, FIREFLY can efficiently explore
different target priors by rerunning only the low-cost importance-sampling
step.
\Refereeone{For completeness, we report the corresponding allocated core-hours and
likelihood-evaluation counts for both methods in
Table~\ref{tab:firefly_computational_cost}. For FIREFLY, the reported
likelihood-evaluation count is the number recorded by \texttt{bilby} during
the auxiliary inference.}

\begin{table}[t]
\centering
\caption{Stepwise runtime of FIREFLY for different numbers of modes $N$ using
the \texttt{nessai} and \texttt{dynesty} samplers, decomposing those reported in Fig.~2 of the main text
into the auxiliary-inference and importance-sampling steps.
The reported values are the mean and standard deviation over five
independent runs.
}
\label{tab:firefly_runtime_modes}
\vspace{0.2cm}
\setlength{\tabcolsep}{10pt}
\begin{tabular}{@{}ccccc@{}}
\toprule
Sampler & Number of Modes & Auxiliary inference $[\mathrm{min}]$ & Importance sampling $[\mathrm{min}]$ & Total $[\mathrm{min}]$ \\
\midrule
\texttt{nessai} & $N=1$ & $2.49\pm0.14$ & $0.302\pm0.006$ & $2.79\pm0.14$ \\
                & $N=2$ & $2.70\pm0.10$ & $0.335\pm0.003$ & $3.03\pm0.09$ \\
                & $N=3$ & $2.85\pm0.19$ & $0.377\pm0.006$ & $3.23\pm0.19$ \\
                & $N=4$ & $2.96\pm0.05$ & $0.401\pm0.005$ & $3.36\pm0.05$ \\
                & $N=5$ & $3.18\pm0.19$ & $0.452\pm0.006$ & $3.64\pm0.19$ \\
                & $N=6$ & $3.68\pm0.72$ & $0.522\pm0.069$ & $4.20\pm0.78$ \\
\midrule
\texttt{dynesty} & $N=1$ & $149\pm2$ & $0.302\pm0.013$ & $149\pm2$ \\
                 & $N=2$ & $158\pm5$ & $0.333\pm0.010$ & $158\pm5$ \\
                 & $N=3$ & $201\pm4$ & $0.371\pm0.014$ & $202\pm4$ \\
\bottomrule
\end{tabular}
\end{table}

\begin{table}[H]
\centering
\caption{Computational costs of FIREFLY and the full-parameter sampling for
different numbers of modes $N$, using the \texttt{nessai} and
\texttt{dynesty} samplers. The
 allocated core-hours are calculated as the reported
wall times multiplied by the 64 allocated CPU cores. The likelihood-evaluation columns give the
numbers recorded by \texttt{bilby} during the full-parameter and auxiliary
sampling, respectively. The reported values are the mean and standard deviation
over five independent runs.}
\label{tab:firefly_computational_cost}
\vspace{0.2cm}
\setlength{\tabcolsep}{8pt}
\begin{tabular}{@{}cccccc@{}}
\toprule
& & \multicolumn{2}{c}{Allocated core-hours}
& \multicolumn{2}{c}{Likelihood evaluations} \\
\cmidrule(lr){3-4}\cmidrule(lr){5-6}
Sampler & Number of Modes
& Full & FIREFLY
& Full $[10^{6}]$ & Auxiliary $[10^{4}]$ \\
\midrule
\texttt{nessai}
& $N=1$ & $4.69\pm0.17$ & $2.98\pm0.15$ & $0.165\pm0.008$ & $7.6\pm0.4$ \\
& $N=2$ & $10.9\pm0.5$ & $3.23\pm0.10$ & $0.39\pm0.02$ & $7.6\pm0.4$ \\
& $N=3$ & $26\pm3$ & $3.4\pm0.2$ & $0.91\pm0.09$ & $7.4\pm0.6$ \\
& $N=4$ & $112\pm20$ & $3.59\pm0.05$ & $4.0\pm0.7$ & $7.32\pm0.03$ \\
& $N=5$ & $420\pm90$ & $3.9\pm0.2$ & $13\pm3$ & $7.3\pm0.5$ \\
& $N=6$ & $870\pm70$ & $4.5\pm0.8$ & $24.6\pm1.8$ & $7.4\pm0.3$ \\
\midrule
\texttt{dynesty}
& $N=1$ & $522\pm6$ & $159.2\pm2.0$ & $2.105\pm0.014$ & $54.2\pm0.4$ \\
& $N=2$ & $1636\pm14$ & $169\pm5$ & $5.74\pm0.03$ & $52.9\pm0.3$ \\
& $N=3$ & $4440\pm60$ & $215\pm5$ & $13.12\pm0.15$ & $52.9\pm0.6$ \\
\bottomrule
\end{tabular}
\end{table}

\section{Additional robustness and scalability tests}
In the simulations presented above, the inference configurations, such as the
priors and mode choices,
are fixed, while the speedup of FIREFLY is expected to be
configuration-dependent to some extent.
In this section, we perform additional tests to explore the performance of
FIREFLY with the \texttt{nessai} sampler.

\subsection{Runtime tests on prior and mode choices}
\label{sec:runtime_prior_mode}

\begin{figure}[H]
  \centering
  \includegraphics[width=0.7\textwidth]{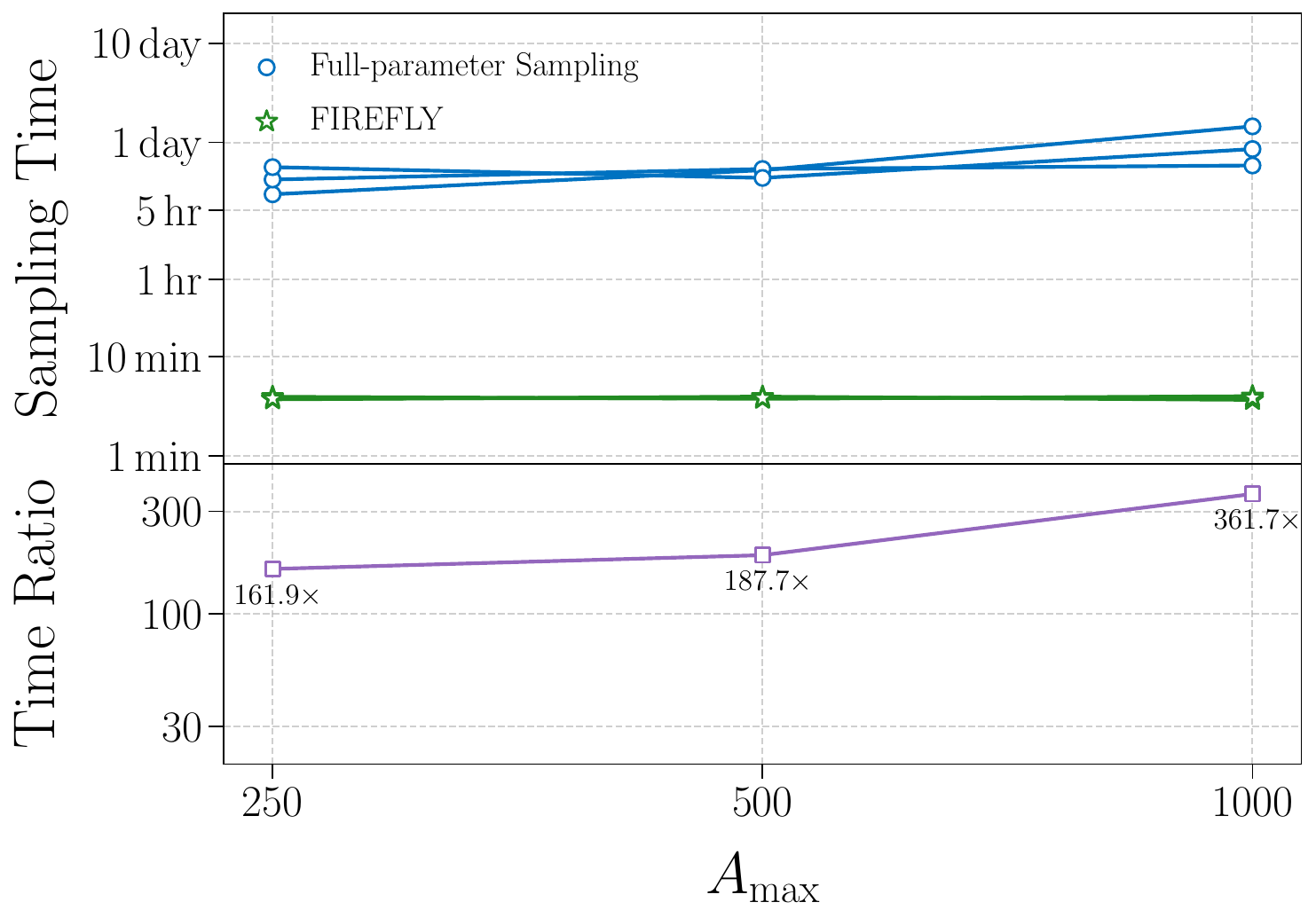}
  \caption{Runtime comparison for varying prior ranges of the amplitude
  parameters, with $A_{\max}$ shown in units of $10^{-20}$.
  For each $A_{\max}$, three independent runs are performed for both
  the full-parameter sampling and FIREFLY. The speed-up factor is calculated as
  the ratio of the runtime averaged over the three runs.}
  \label{fig:time_changeA}
\end{figure}

We first test the dependence on the target priors of the mode amplitudes.
As discussed in Sec.~\ref{sec:Quantitative metrics}, an analysis involving
multiple prior choices can be efficiently investigated
by the importance-sampling step in FIREFLY, while the auxiliary inference only
needs to be performed once. Therefore, we only compare the runtime of
FIREFLY and the full-parameter sampling for cases in which
only a single target prior $\pi_{\otimes}$ is considered.
In the main text, we choose flat priors of $A_{\ell mn}\sim {\cal U}(0,500\times
10^{-20})$, while here we vary the upper bound
$A_{\max}$ of the uniform amplitude priors. The timing results are shown in
Fig.~\ref{fig:time_changeA}. When $A_{\max}$ is increased from
$250\times10^{-20}$ to $1000\times10^{-20}$, the runtime of the full-parameter
sampling increases, consistent with the fact that nested sampling needs to
shrink from a larger prior volume to the posterior peak. In contrast,
the FIREFLY runtime remains around four minutes, since the auxiliary inference
is performed with a fixed flat prior in $\bmB$ that is independent of the target
prior, and the computational cost of evaluating the importance weights in the importance-sampling step is
also weakly affected by the change of the target prior. The speedup factor of FIREFLY over the full-parameter sampling therefore
increases from about $160$ to $360$, providing more than two orders of magnitude
acceleration.

\begin{figure}[H]
  \centering
  \includegraphics[width=0.7\textwidth]{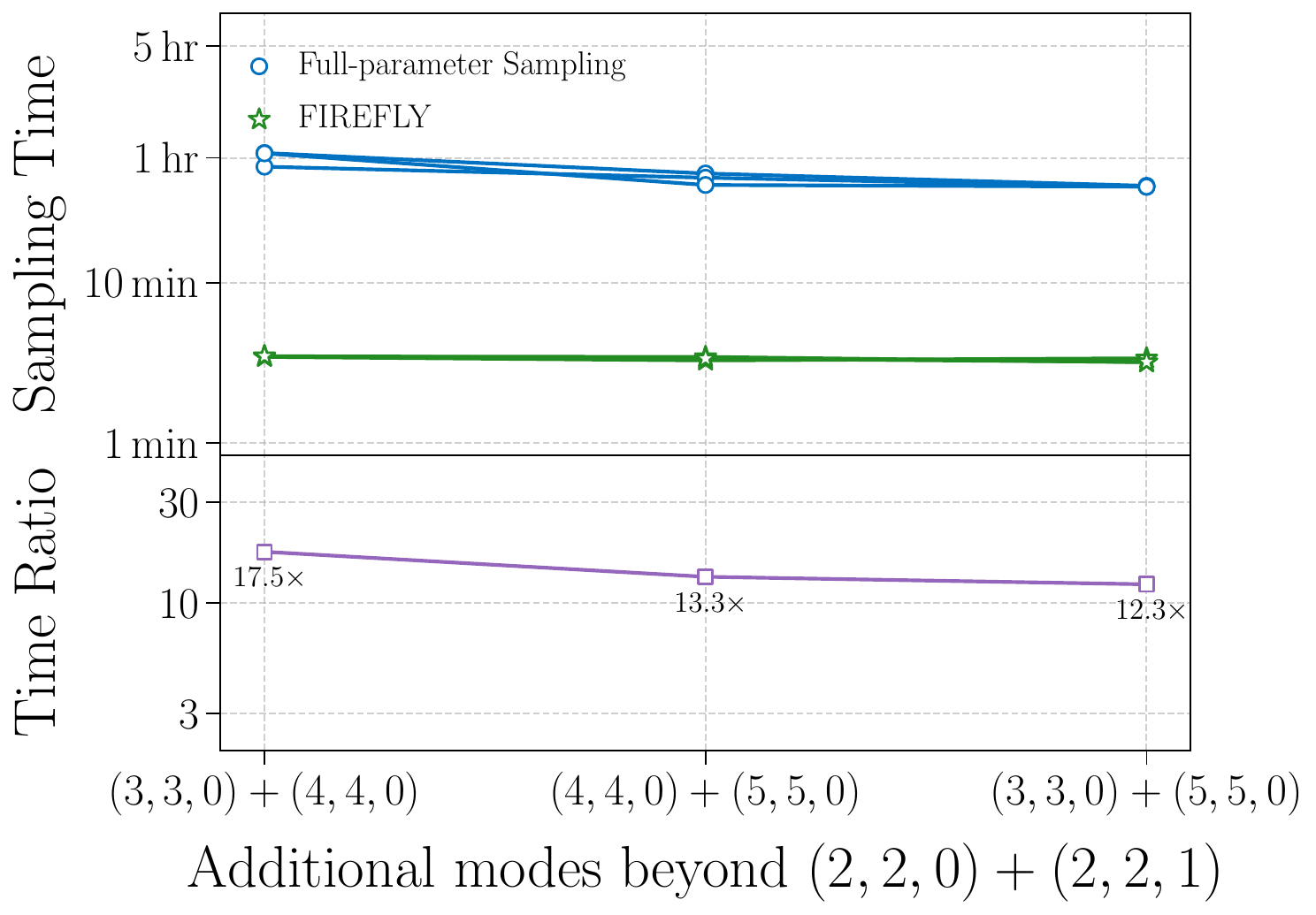}
  \caption{Runtime comparison for varying mode configurations. The dominant
  $(2,2,0)$ and $(2,2,1)$ modes are kept fixed, while the other two modes are
  replaced by different combinations of subdominant modes.}
  \label{fig:time_changemodes}
\end{figure}

We next test the dependence on the mode choice. In the first set of tests, we
fix the number of modes to be $N=4$ and compare different mode combinations.
The dominant $(2,2,0)$ and $(2,2,1)$ modes are kept fixed, while the other two
modes are chosen as $(3,3,0)+(4,4,0)$, $(4,4,0)+(5,5,0)$, and
$(3,3,0)+(5,5,0)$, respectively. As shown in
Fig.~\ref{fig:time_changemodes}, the FIREFLY runtime varies only mildly, from
about $3.2$ to $3.4$ minutes, consistent with the $N=4$ runtime in
Table~\ref{tab:firefly_runtime_modes}. The runtime of the full-parameter
sampling shows a stronger dependence on the specific mode combination. This
trend is qualitatively correlated with the individual mode SNRs listed in
Table~\ref{tab:qnm_injection_tdiA}: combinations involving stronger
subdominant modes tend to require longer full-parameter sampling times.  The mode configuration in
the main text, $(2,2,0)+(2,2,1)+(3,3,0)+(3,3,1)$, contains the $(3,3,1)$ mode
with an even higher SNR, and correspondingly has a longer full-parameter
sampling time. Therefore, the speedup of FIREFLY also depends on the mode choice through the
runtime of the full-parameter sampling. Nevertheless, we note that the speedup factors for 
four-mode analysis, $\sim 10$--$20$ for the three combinations in this section
and $\sim 30$ in the main text, all lie between the speed-ups of the $N=3$ 
and
$N=5$ cases (see Fig.~2 of the main text), so the overall trend that FIREFLY
delivers more acceleration for larger $N$ remains valid.

In the second mode-choice test, we keep a six-mode configuration but replace the
$(3,3,1)$ mode by the $(2,1,0)$ mode, while leaving the other modes unchanged.
For this configuration, the full-parameter sampling time increases to about one
day on average over three runs
, whereas the FIREFLY runtime remains around four minutes, corresponding to a
speed-up close to $400$. This increase in the full-parameter runtime is likely
related to the similarity between the frequencies and damping times of the
$(2,1,0)$ and $(2,2,0)$ modes, which can make the two modes harder to resolve
and complicate the sampling.
In contrast, FIREFLY is less affected by this mode confusion because the
amplitude parameters are marginalized in the auxiliary-inference step.

For all additional simulations discussed above, FIREFLY and the
full-parameter sampling give consistent posterior distributions. 
Overall, these tests show that the exact speed-up factor is configuration
dependent, and better to be understood as an order-of-magnitude estimate. At the same time, we find that
the computational time of FIREFLY is only mildly affected across the configurations tested here, staying at the level of a few
minutes. This supports the robustness of FIREFLY as
a fast inference strategy for multimode ringdown analyses.

\subsection{Eight-mode stress test}

As a more demanding test of FIREFLY in a higher-dimensional setting, we further
perform an eight-mode analysis by including the $(2,1,0)$ and $(2,1,1)$ modes in
addition to the six modes used in the main text, with all other settings
unchanged. For this configuration, the full-parameter sampling with
\texttt{nessai} did not converge within five days, with \texttt{dlogz} still
larger than $1000$. In contrast, FIREFLY produced stable posterior samples
within a few minutes, demonstrating its practical capability in this more
challenging multimode case.

\begin{figure}[H]
  \centering
  \includegraphics[width=0.95\textwidth]{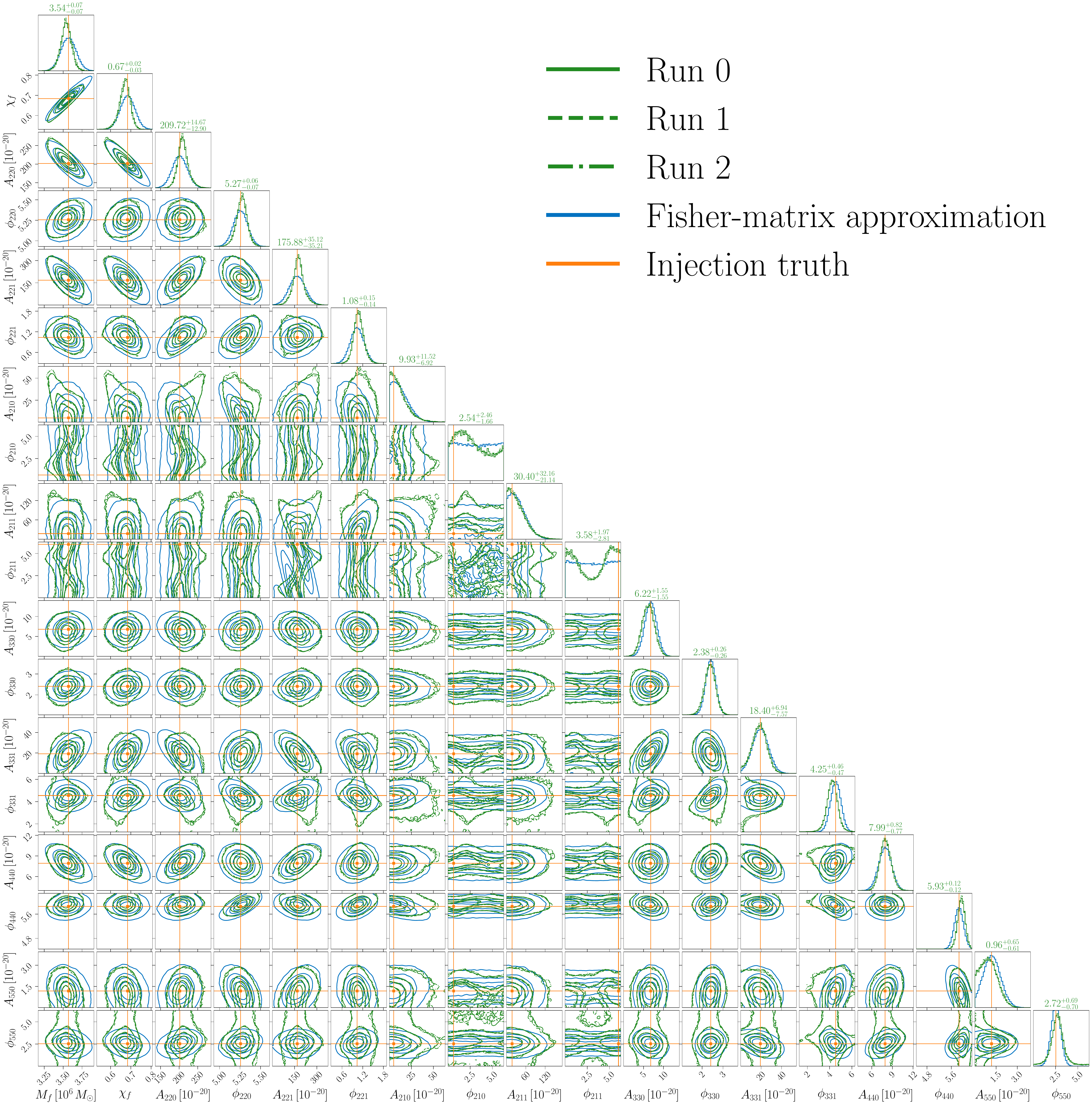}
  \caption{Posteriors for the eight-mode analysis with additional modes
  $(2,1,0)$ and $(2,1,1)$, obtained with three independent runs of FIREFLY
  (plotted in different line styles). The blue lines represent the posterior
  obtained with the Fisher-matrix approximation.}
  \label{fig:SM_corner_8modes}
\end{figure}

Figure~\ref{fig:SM_corner_8modes} shows the posteriors obtained from three
independent FIREFLY runs for the zero-noise injection. The three runs are
mutually consistent and compatible with the injected values, indicating the
stability of the sampling. Since a direct comparison with full-parameter
sampling is not available in our test, we use the Fisher-matrix
approximation as an independent sanity check, where the likelihood is
approximated as a Gaussian function. The FIREFLY
contours are consistent with the Fisher-matrix estimate in scale and orientation,
while also capturing non-Gaussian features beyond the Fisher approximation. This
provides indirect evidence supporting the reliability of FIREFLY in this
eight-mode stress test.

\begin{figure}[H]
  \centering
  \includegraphics[width=0.9\textwidth]{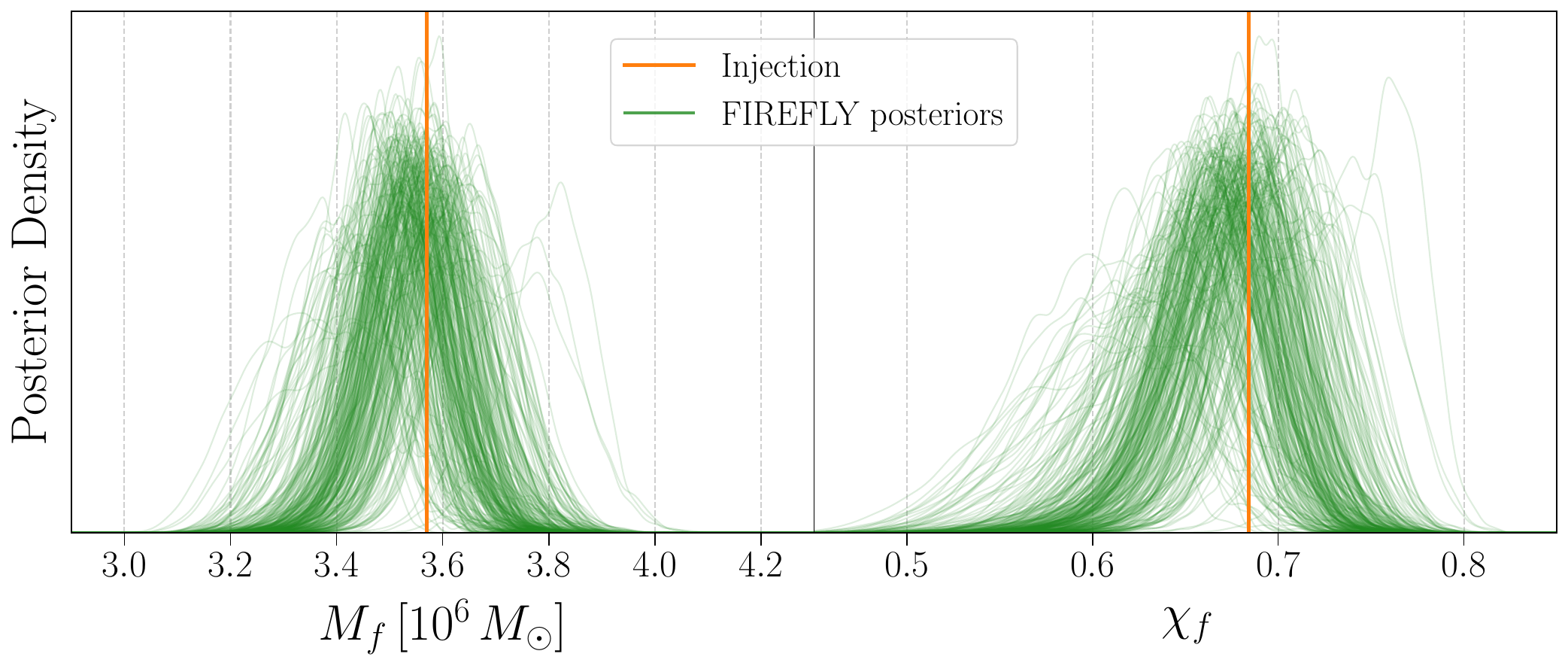}
  \caption{Marginalized posteriors of the final mass $M_f$ and spin $\chi_f$ for
  300 different noise realizations, obtained with FIREFLY in the eight-mode
  analysis. The orange lines represent the injected values.}
  \label{fig:SM_posteriors_noise_realizations}
\end{figure}

\begin{figure}[H]
  \centering
  \includegraphics[width=0.7\textwidth]{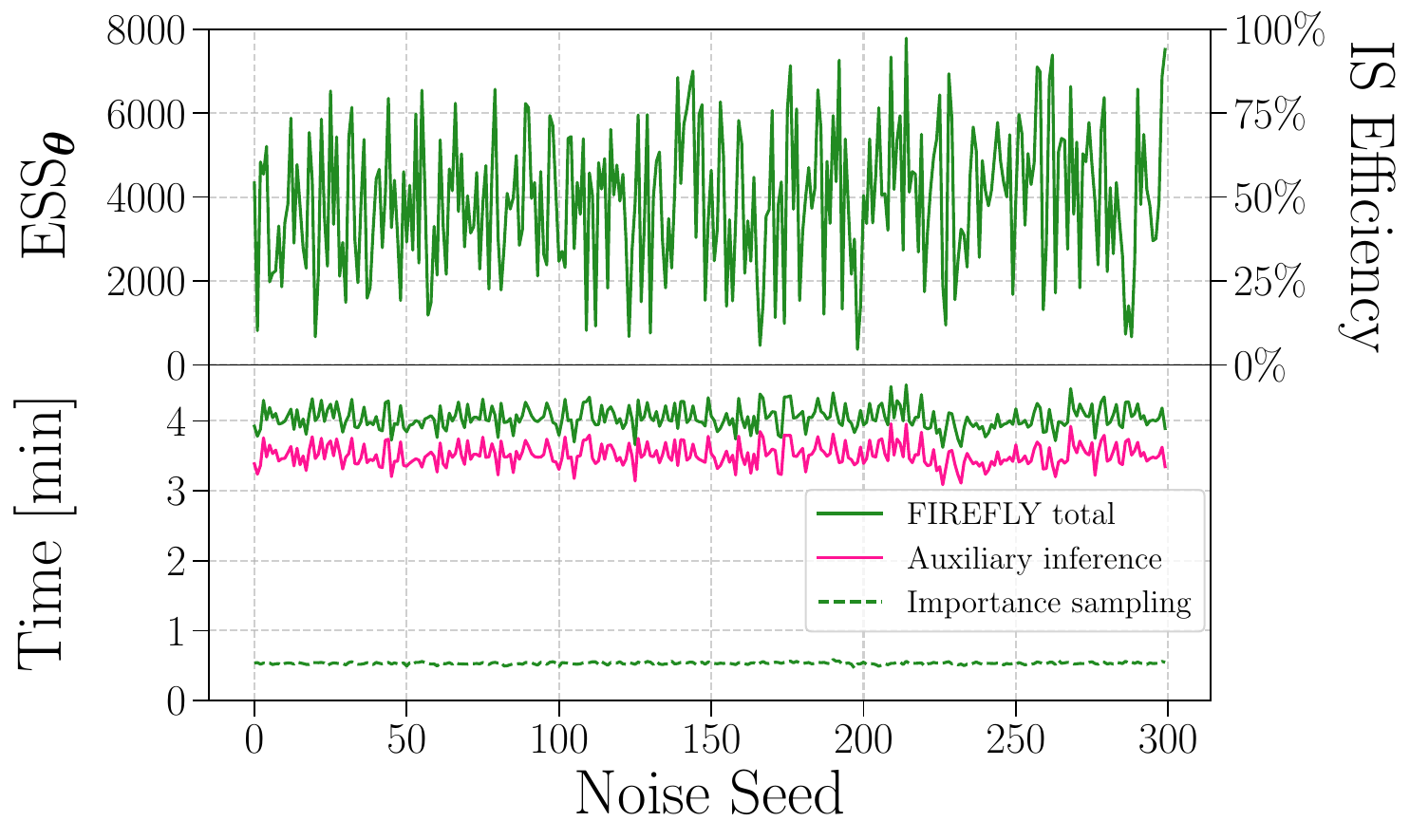}
  \caption{ESS and runtime of FIREFLY for 300 different noise realizations in the
  eight-mode analysis. The IS efficiency in the right $y$-axis of the upper panel
  is defined as the ratio of the effective sample size to $8000$, since the
  sample size in the auxiliary inference is $8005\pm205$ over the 300 noise
  realizations.}
  \label{fig:SM_noise_ESS_time_vs_seed}
\end{figure}

To further test the computational performance stability of FIREFLY, we repeat
the eight-mode analysis over 300 different noise realizations. In
Fig.~\ref{fig:SM_posteriors_noise_realizations} we show the marginalized
posteriors of the final mass $M_f$ and spin $\chi_f$ for these noise
realizations, which surround the injected values in a stable manner. In
Fig.~\ref{fig:SM_noise_ESS_time_vs_seed}, we show the effective sample size
(ESS) and runtime for these noise realizations. Over the 300 noise realizations,
the ESS has a mean value of around $4000$ with a standard deviation of around
$1600$, and only 7 ($2.3\%$) realizations have ESS smaller than $800$. For the
runtime, the total runtime of FIREFLY is stable around $4$ minutes, with the main
variation coming from the auxiliary inference step and a standard deviation of
less than $10$ seconds. Therefore, FIREFLY's performance is stable across
different noise realizations, and the eight-mode analysis can be completed within
a few minutes.

\Refereeone{For representative low-ESS cases, we find a substantial mismatch
between the auxiliary and target posteriors in $\bmt$ space, concentrating the
marginal weights on a small subset of samples. This may arise because the flat
auxiliary prior favours larger amplitudes, coupled with their correlations
with mass and spin, and it can become more pronounced with multiple weak modes.
Increasing the MC samples per marginal weight to $n_w=2\times10^5$ and the
conditional candidates to $n_{\rm B}=2\times10^4$ produces
little change in the ESS or recovered posteriors, indicating that MC
fluctuations are not the main cause of the low ESS in these cases.}

\begin{figure}[!t]
\centering
\includegraphics[width=0.95\textwidth]{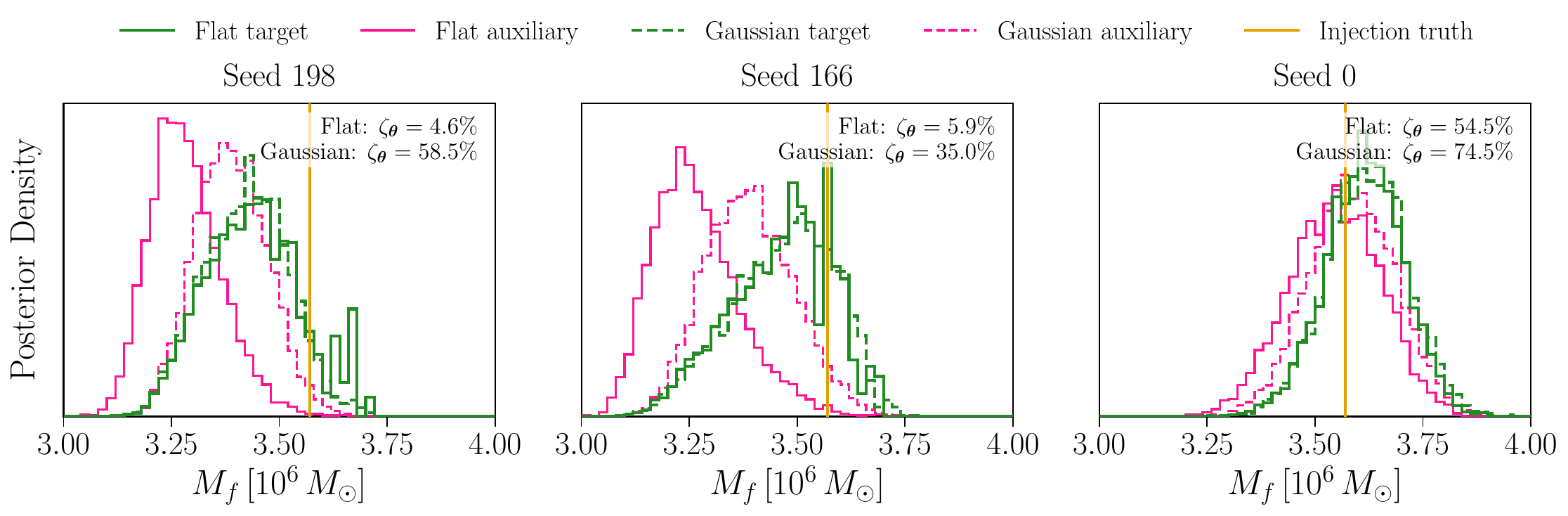}
\caption{\Refereeone{Auxiliary and target marginal posterior distributions of
the final mass under flat and Gaussian auxiliary priors,
for seeds 198, 166, and 0 (left to
right). Note that in each panel,
the green lines indicate the same target posterior calculated via flat (solid)
and Gaussian (dashed) auxiliary priors.}}
\label{fig:SM_gaussian_mass}
\end{figure}

\Refereeone{We therefore test an isotropic
Gaussian auxiliary prior
$\pi(\bmB|\bmt,\apc_{\rm G})=\mathcal N(\bmB;0,\sigma^2 I_{2N})$,
where $I_{2N}$ is the $2N\times 2N$ identity matrix, and $\sigma=A_{\max}/3$ with $A_{\max}$ the maximum amplitude in the target
prior. This suppresses larger amplitudes while preserving
FIREFLY's analytical marginalization and efficient conditional sampling.
Table~\ref{tab:SM_auxiliary_refinement} compares the flat and Gaussian auxiliary
priors for noise realizations 198, 166, and 0. The first two seeds have low ESS with
the flat auxiliary prior, while seed 0 has an ESS near the average ESS over 300 noise realizations. 
 For the two lowest-ESS cases,
$\zeta_{\bmt}=\mathrm{ESS}_{\bmt}/n_{\bmt}$ increases from $4.6\%$ and $5.9\%$
to $58.5\%$ and $35.0\%$, respectively, where $n_{\bmt}$ is the number of
auxiliary $\bmt$ samples used for reweighting. We also compute the
Pareto-smoothed importance sampling (PSIS) tail diagnostic $\hat
k$~\cite{JMLR:v25:19-556} from the marginal $\bmt$ weights, which decreases from
$0.60$ and $0.56$ to $0.25$ and $-0.29$, respectively, indicating reduced
auxiliary--target posterior mismatch.
The recovered marginal posterior
distributions are also smoother, shown in Fig.~\ref{fig:SM_gaussian_mass}. In addition, the Gaussian auxiliary prior has
 comparable likelihood-evaluation counts of
approximately $7\times10^4$ compared to the flat auxiliary prior.
}

\begin{table}[!t]
\centering
\small
\caption{\Refereeone{Marginal effective sample fractions $\zeta_{\bmt}$ and
PSIS $\hat k$ for the three configurations. The
$n_w=2\times10^5$ row gives arithmetic means over three independent MC
repetitions; the other rows report single runs.}}
\label{tab:SM_auxiliary_refinement}
\vspace{0.2cm}
\renewcommand{\arraystretch}{1.15}
\setlength{\tabcolsep}{6pt}
\Refereeone{\begin{tabular}{@{}lcccccc@{}}
\toprule
& \multicolumn{2}{c}{Seed 198} & \multicolumn{2}{c}{Seed 166} & \multicolumn{2}{c}{Seed 0}\\
\cmidrule(lr){2-3}\cmidrule(lr){4-5}\cmidrule(lr){6-7}
Configuration & $\zeta_{\bmt}$ & $\hat k$ & $\zeta_{\bmt}$ & $\hat k$ & $\zeta_{\bmt}$ & $\hat k$\\
\midrule
Flat, $n_w=5\times10^4$ & $4.6\%$ & $0.60$ & $5.9\%$ & $0.56$ & $54.5\%$ & $-0.18$\\
\addlinespace
Flat, $n_w=2\times10^5$ & $4.2\%$ & $0.59$ & $6.0\%$ & $0.54$ & $54.4\%$ & $-0.30$\\
\addlinespace
Gaussian, $n_w=5\times10^4$ & $58.5\%$ & $0.25$ & $35.0\%$ & $-0.29$ & $74.5\%$ & $0.08$\\
\bottomrule
\end{tabular}}
\end{table}

\Refereeone{We recommend warnings for $\zeta_{\bmt}<10\%$ and for a PSIS tail diagnostic $\hat k>0.5$,
computed from the marginal $\bmt$ weights, followed by auxiliary-prior
adjustment if needed. The $\hat k>0.5$ threshold is a conservative choice
motivated by the risks of second-moment in unsmoothed importance weights.
In the second IS step, our tests show that the current choice of
$n_{\rm B}=5000$ conditional samples per $\bmt_j$ is sufficient for stable
recovery of the conditional $\bmB$ posteriors. Nevertheless, the conditional $\bmB$ weights can still have a divergent
second moment near zero amplitude, which may be mitigated by mixture proposals.
A more comprehensive investigation of the auxiliary-prior choice, mixture
proposals, or adaptive sampling strategies is left for future work.}

\clearpage
\bibliographystyle{apsrev}
\bibliography{refs}

\end{document}